\documentclass[12pt,twoside]{article}
\pdfoutput=1

\newlength{\dinwidth}
\newlength{\dinmargin}
\usepackage{amsmath, amsthm, amsfonts, amssymb, bm, mathrsfs, indentfirst,
sectsty, graphicx, fancyhdr, slashed, fullpage, color, authblk, easybmat}
\usepackage{hyperref}

\begin{document}
%\maketitle

\begin{titlepage}

\flushright{IFT-UAM/CSIC-14-070}\\[2cm]
\begin{center}
 {\Large \bf \sc Anomalous Magneto Response and the St\"uckelberg Axion in
Holography}
\\[1.52cm]
{\large Amadeo Jimenez-Alba\footnote{amadeo.j@gmail.com},
Karl Landsteiner\footnote{karl.landsteiner@csic.es} and Luis
Melgar\footnote{luis.melgar@csic.es}  }

\bigskip

{}{\small \it Instituto de F\'\i sica Te\'orica IFT-UAM/CSIC, Universidad
Aut\'onoma de Madrid,
28049 Cantoblanco, Spain}\\
\end{center}

\bigskip

\begin{abstract}
We study the magneto response with non-conserved currents in Holography.
Non-conserved currents are dual
to massive vector fields in AdS. We introduce the mass in a gauge invariant way
via the St{\" u}ckelberg
mechanism. In particular we find generalizations of the Chiral Magnetic Effect,
the Chiral Separation Effect and the
Chiral Magnetic Wave. Since the associated charge is not conserved we need to
source it explicitly by a coupling,
the generalization of the chemical potential. In this setup we find that in
general the anomalous transport phenomena 
are still realized. The values we find
for non-zero mass connect however continuously to the values of the anomalous
conductivities of the {\bf \em consistent} currents. 
i.e. the  proper chiral magnetic effect vanishes for all masses (as it does for
the consistent current
in the zero mass case) whereas the chiral separation effect is fully present. 
The generalization of the chiral magnetic wave shows that for small momenta
there is no propagating wave but
two purely absorptive modes (one of them diffusive). At higher momenta we
recover the chiral magnetic wave
as a combination of the two absorptive modes. 
We also study the negative magneto resistivity and find that it grows
quadratically with the magnetic field. The chiral magnetic wave and the negative magneto resistivity are manifestations
of the chiral magnetic effect that
takes place when the (non-conserved) charge is allowed to fluctuate freely in
contrast to the case where the charge
is fixed by an explicit source. 
Since the (classical) $U(1)_A$ symmetry of QCD 
is not at all a symmetry at the quantum level we also argue that using
massive vectors in AdS to model the axial
singlet current might result in a more realistic holographic model of QCD and
should be a good starting point to investigate 
the dynamics of anomalous transport in the strongly coupled quark gluon plasma. 
\end{abstract}
\end{titlepage}

%%%%%%%%%%%%%%%%%%         INTRO                   %%%%%%%%%%%%%%%%%%%%%%%%%

\section{Introduction}
Anomalies in the quantum theories of chiral fermions belong to the most
emblematic properties of relativistic quantum field theory
\cite{Adler:1969gk,Bell:1969ts} (see \cite{Bertlmann:1996xk,Bilal:2008qx} for
reviews). They provide stringent consistency conditions on possible gauge
interactions but also predict physical processes that would be otherwise highly
suppressed such as the decay of the neutral pion into two photons.\\ 
Anomalies are not only important for the phenomenology of particle physics but
they also are of utmost importance to the theory of quantum many body systems
containing chiral fermions. Anomaly cancellation plays a crucial role in the
field theoretic understanding of the electro response of quantum hall fluids for
example. Chiral fermions appear as edge states and the associated anomalies have
to be canceled by appropriate anomaly inflow from a gapped bulk reservoir of
charge. \\
More recently the focus has been on ungapped chiral bulk fermions that give rise
to new anomaly related transport phenomena in the presence of a magnetic field
(chiral magnetic effect \cite{Fukushima:2008xe}) and/or vortices (chiral
vortical effect \cite{Erdmenger:2008rm, Banerjee:2008th}). 
The chiral magnetic effect describes the generation of a current in the presence
of a magnetic field
\begin{equation}
 \vec{J} = \sigma_B\vec{B}\,.
\end{equation}
The associated chiral magnetic conductivity \cite{Kharzeev:2009pj} can be
calculated from first principles via a Kubo type formula
\begin{equation}
\label{eq:Kubo}
 \sigma_{B} = \lim_{k_z\rightarrow 0}\frac{i}{k_z} \langle J_x J_y \rangle
(\omega=0,\vec{k}=k_z\hat e_z)\,.
\end{equation}
Since these effects owe their existence to the presence of (global) anomalies
one could expect that their values are universal and independent from
interactions. Indeed calculations with free fermions
\cite{Landsteiner:2011cp,Loganayagam:2012pz} give the same result as infinitely
strongly coupled theories defined via the AdS/CFT correspondence
\cite{Newman:2005hd, Landsteiner:2011iq}. Furthermore it was shown that the
anomalous conductivities are completely determined in hydrodynamics or in
effective action approaches \cite{Son:2009tf, Neiman:2010zi, Banerjee:2012iz}
(with the exception of the gravitational anomaly contribution, whose model
independent determination needs additional geometric arguments
\cite{Jensen:2012kj}). Therefore the values of the chiral conductivities related
to purely global anomalies are subject to a non-renormalization 
theorem akin to the Adler-Bardeen theorem \cite{Golkar:2012kb}. 

Chiral conductivities do get renormalized however in the case when the gauge
fields appearing in the anomalous divergence of the current are dynamical
\cite{Hou:2012xg,Golkar:2012kb}. An example is the singlet $U(1)_A$ current in
QCD. Its anomaly is of the form
\begin{equation}
 \partial_\mu J^\mu_A = \epsilon^{\alpha\beta\gamma\delta} \left( \frac{N_c
\sum_f q_f^2}{32\pi^2} F_{\alpha\beta}F_{\gamma\delta}
+ \frac{N_f}{16\pi^2} \mathrm{tr} ( G_{\alpha\beta} G_{\gamma\delta} ) +
\frac{N_c N_f}{96\pi^2} F^5_{\alpha\beta}F^5_{\gamma\delta}
\right)
\end{equation}
Here $F$ is the electromagnetic field strength, $G$ the gluon field strength and
$F^5$ the field strength of an axial gauge field $A^5_\mu$ whose only purpose is
to sum up insertions of the axial current in correlation functions, i.e. there
is no associated kinetic term. $N_c$ and $N_f$ are the numbers of colors and
flavors respectively. In this case it has first been shown in \cite{Hou:2012xg}
that the vortical conductivity receives two loop corrections whereas later on is
has been argued in an effective field theory approach that all chiral
conductivities  receive higher loop corrections once dynamical gauge fields
enter the anomaly equation \cite{Jensen:2013vta}. 

It has been argued long ago by 't Hooft that in such a situation one should not
think of the classically present $U(1)_A$ symmetry as a  fact symmetry at all
on the quantum level \cite{'tHooft:1986nc}. In fact in asymptotically free
theories such as QCD there might survive only a discrete subgroup because of
instanton contributions. This discrete subgroup can be further broken
spontaneously via chiral symmetry breaking but since it was not a symmetry to
begin with there is also no associated Goldstone boson, which explains the high
mass of the $\eta'$ meson in QCD. A related fact is that the corresponding
triangle diagram receives higher loop corrections via photon-photon or
gluon-gluon rescattering. These higher order diagrams lead to a non-vanishing
anomalous dimension for the axial current operator $J^\mu_A$. See 
\cite{Adler:2004qt,Ioffe:2006ww} for recent reviews.\\

These considerations motivate us to study the anomalous magneto response of
massive vector fields in holography. Our philosophy is as follows. In quantum
field theory we would have to study the path integral
\begin{equation}\label{eq:pathintegral}
 Z = \int D\Psi D\bar\Psi D\mathcal{A}_q \, \exp\left[ i \int d^4x \left( -\frac
1 2 \mathrm{tr} (G.G) + \bar\Psi D \Psi + \theta \mathcal{O}_A
  \right)\right]\,,
\end{equation}
where $A_q$ stands collectively for the dynamical gauge fields, $G$ is their
field strength tensor and $\mathcal{O}_A$ is the (operator valued) anomaly
\begin{equation}
 \mathcal{O}_A = \epsilon^{\alpha\beta\gamma\delta}\left(\frac{N_f}{16\pi^2}
\mathrm{tr} ( G_{\alpha\beta} G_{\gamma\delta} ) + \frac{N_c N_f}{96\pi^2}
F_{\alpha\beta}F_{\gamma\delta}\right)\,.
\end{equation}
Since the anomaly is a quantum operator we need to define a path integral that
allows to calculate correlations functions of this anomaly operator. This means
that we need to introduce the source field $\theta(x)$ coupling to
$\mathcal{O}_A$. For the same reason we also have to include a source for the
anomalous current $J^\mu$. This source is the non-dynamical gauge field which
from now on we denote by $A_\mu$. The covariant derivative in
(\ref{eq:pathintegral}) contains both, the dynamical $A^\mu_q$ and the
non-dynamical gauge fields.
If we define the effective action $\exp(i W_{\mathrm{eff}}[A,\theta] ) = Z$ it
is basically guaranteed by construction that this effective action enjoys the
gauge symmetry
\begin{equation}\label{eq:gaugesymmetry}
 \delta A_\mu = \partial_\mu \lambda \;,~~~~~~~ \delta \theta = - \lambda
\;,~~~~~~~ \delta W_\mathrm{eff} =0\,.
\end{equation}
We now replace the (strongly coupled) dynamics of the gluon (and fermion)
fields, i.e. the path integral over $A_q$, $\Psi$ and $\bar\Psi$ by the dynamics
of classical fields propagating in Anti-de Sitter space. The gravity dual should
allow to construct $W_{eff}[A,\theta]$ as the on-shell action of a field theory
in Anti-de Sitter space containing a vector field $A_\mu$ and a scalar $\theta$
obeying the gauge symmetry (\ref{eq:gaugesymmetry}). In addition, as we have
argued before, the vector field should source a non-conserved current. Since
Anti-de Sitter space is related to the theory having an additional conformal
symmetry then the four-dimensional current is non-conserved if and only if its
dimension is different from three. This in turn means that the bulk vector field
in our AdS theory has to be a massive vector and it is precisely the gauge
symmetry (\ref{eq:gaugesymmetry}) that allows the inclusion of a gauge invariant
St\"uckelberg mass in the bulk AdS theory. The anomaly also includes the global
part proportional to the field strengths of the non-dynamical gauge field
therefore we also need to include a five dimensional Chern-Simons term in our
AdS dual. The relation of the St\"uckelberg field in holography to the anomaly
has been first pointed out in \cite{Klebanov:2002gr} and the necessity to
include it in holographic studies of the anomalous transport has very recently
also been emphasized in \cite{Gursoy:2014ela}.\\
Moreover, since we have application to the physics of the strongly coupled quark
gluon plasma in back of our head, we are lead to study a massive St\"uckelberg
theory with Chern-Simons term at high temperature, i.e. in the background of an
AdS black brane. We make however one more simplifying assumption. We do not
study any correlations functions including the energy momentum tensor. Therefore
we can resort to the so called probe limit in which we ignore the back reaction
for the gauge field theory onto the geometry. \\

The paper is organized as follows. In section \ref{sec:U1model} we define a
simple model with one massive vector field. We calculate the (holographically)
normalized non-conserved current and compare to the massless case. Then we
study the generalization of the chiral magnetic conductivity defined via the
Kubo formula (\ref{eq:Kubo}). We find that the chiral conductivity still exists
and in terms of an appropriately defined dimensionless number gets even enhanced
compared to the massless case. In the limit of vanishing mass we recover the
value of the chiral magnetic conductivity in the {\em \bf consistent current}.
As is well-known this is $2/3$ of the standard value most commonly cited (which
corresponds to the covariant definition of the current). \\
We remind the reader of the fact that the chiral magnetic effect 
in the consistent current for the $U(1)^3$ 
anomaly of a single Weyl fermion takes the form
\begin{equation}\label{eq:cmeu1}
 \vec{J} = \left(\frac{\mu}{4\pi^2} - \frac{A_0}{12\pi^2} \right)\vec{B}
\end{equation}
whereas for the $AVV$ anomaly of a single Dirac fermion with a vector current 
preserving regularization it is
\begin{equation}\label{eq:cmeVAA}
 \vec{J}_V = \left(\frac{\mu_5}{2\pi^2} - \frac{A_0^5}{2\pi^2} \right)\vec{B}
\end{equation}
with $\mu$, $\mu_5$ the (axial) chemical potentials and $A_0$ and $A_0^5$ the
background values of the (axial) gauge fields that do not necessarily coincide
with the chemical potentials. The customary gauge choice $A_0=\mu$ and $A_0^5=\mu_5$
leads to the factor $2/3$ in the $U(1)^3$ case and to a vanishing CME in the
$AVV$ case\footnote{If an axion background is present there is also a term 
proportional to $\partial_t \theta \vec B$. We also emphasize that the anomaly
makes the (axial) ``gauge'' field an observable precisely via the terms
in (\ref{eq:cmeu1}), (\ref{eq:cmeVAA}). See e.g. the discussion in 
\cite{Landsteiner:2012kd}.} If one expresses the CME however in terms of the covariant
currents the terms depending on the gauge fields are absent. Finally we note that
the relation between covariant and consistent currents are 
$J^\mu_{\mathrm{cov}} = J^\mu_\mathrm{cons} + \frac{1}{24\pi^2} \epsilon^{\mu\nu\rho\lambda}
A_\nu F_{\rho\lambda}$ for the $U(1)^3$ anomaly and 
$J^\mu_{\mathrm{cov},V} = J^\mu_\mathrm{cons,V} + \frac{1}{12\pi^2} \epsilon^{\mu\nu\rho\lambda}
A^5_\nu F_{\rho\lambda}$. In these expressions the currents and Chern Simons terms
all have dimension three.

In section \ref{sec:U1U1model} we consider a massive and a massless vector field
in the bulk. Our motivation is that the proper chiral magnetic effect stems from
an interplay of vector- and axial symmetries. The vector symmetry can be taken
as the usual electromagnetic $U(1)$. While the electromagnetic gauge fields are
still quantum operators we can assume in the quark gluon plasma context that
electromagnetic interactions are weak and to first approximation we might model
the vector $U(1)$ as a non-dynamical gauge field. Furthermore the vector current
of electromagnetic interactions has to be exactly conserved. We compute the
chiral magnetic conductivity and the conductivity related to the chiral
separation effect. We find that the chiral separation effect is fully realized
whereas the chiral magnetic conductivity vanishes. Again we point out that these
are the same results that hold for the consistent currents in the case when
also the axial current is modeled by a massless vector field. Then we study the
chiral magnetic wave \cite{Kharzeev:2010gd} and compare our findings to a simple
hydrodynamic model in which we include a decay width for the axial charge by
hand. We find basically a perfect match between the modes of the
phenomenological model and the low lying quasinormal modes of the holographic
model. For small momenta we find absence of a propagating wave, whereas for
large enough momentum there is indeed a propagating (damped) wave which is the
generalization of the chiral magnetic wave.
Finally we also study the negative magneto-resistivity induced by the anomaly
in a constant magnetic field background. We find by numerical analysis that the
negative magneto-resistivity depends quadratically on the magnetic field. The
optical conductivity has a Drude peak form whose height is determined by the
inverse of the bulk mass. For large magnetic field a gap opens up in the optical
conductivity and we also check that the spectral weight gets shifted from the
gap region into the peak region such that  a sum rule of the form $d \left(\int
d\omega \sigma(\omega)\right) /dB =0$ holds.\\ We present our conclusions in section
\ref{sec:conclusions}, summarize and discuss our results and give some outlook
to possible further generalizations of models with holographic St\"uckelberg
axions.

%%%%%%%%%%             The U(1) Model           %%%%%%%%%%%%%%%%%%%%%%%%%%%%%%

%\newpage
\section{Holographic St\"uckelberg mechanism with a U(1) Gauge Field}
\label{sec:U1model}
In this section we consider Maxwell-Chern-Simons theory in the bulk and give a
mass to the gauge field via St\"uckelberg mechanism
\begin{align}
\label{eq:1}
S=\int d^5x \sqrt{-g}\left(
-\frac{1}{4}F_{\mu\nu}F^{\mu\nu}-\frac{m^2}{2}(A_\mu-\partial_\mu
\theta)(A^\mu-\partial^\mu
\theta)+\frac{\kappa}{3}\epsilon^{\mu\alpha\beta\gamma\delta}(A_\mu
-\partial_\mu \theta)F_{\alpha\beta}F_{\gamma\delta} \right)
\end{align}
The above model provides  a  mass for the gauge field in a consistent
gauge-invariant way. St\"uckelberg terms indeed arise as the holographic
realization of dynamical anomalies, as pointed out for the first time in
\cite{Klebanov:2002gr} (see also \cite{Casero:2007ae} for similar conclusions in
the context of AdS/QCD). This has been also emphasized quite recently by the
authors of \cite{Gursoy:2014ela} for a class of non-conformal holographic
models.\\
As it is well-known, in holography we do not have access to the gauge sector.
This implies that, whereas the global anomaly is implemented by an explicit
Chern-Simons term, the dynamical contribution to the divergence of the current
enters as a mass-term for the gauge field, which induces an explicit
non-conservation. This fits the general expectation that the dynamical anomaly
cannot be switched off because it is not simply a given by a functional of
external fields.\\
Let us also comment on a crucial difference between model (\ref{eq:1}) and
models of holographic superconductors.
Holographic superconductors \cite{Hartnoll:2008kx} also give a bulk mass term to
the gauge field and they might be written in St\"uckelberg form as well
\cite{Franco:2009yz}.
The difference is that the Higgs  mechanism in the bulk uses a massive scalar
field that decays at the boundary and does therefore not
change the asymptotic behavior of the gauge field.
In our case the mass is constant in the bulk and does therefore change the
asymptotic behavior of the vector field as one approaches the
boundary of AdS. \\   

We will work in the probe limit with Schwarzschild-$AdS_5$ as background metric
\begin{equation}
ds^2=-f(r)dt^2+\frac{dr^2}{f(r)}+\frac{r^2}{L^2}\left(dx^2+dy^2+dz^2  \right)\,;
\hspace{1.cm}f(r)=\frac{r^2}{L^2}-\frac{r_H^4}{r^2}\,.
\end{equation}
As usual we make use of rescaling invariance of the theory to set $r_H=1$ and
$L=1$, and therefore $\pi T=1$.\\
The equations of motion are
\begin{align}
\nabla_\nu F^{\nu \mu} -m^2 (A^\mu-\partial^\mu \theta)+ \kappa \epsilon^{\mu
\alpha \beta \gamma \rho} F_{\alpha \beta} F_{\gamma \rho}=0\,,\\
 \label{eq:scalareq}\nabla_\mu \left( A^\mu -\partial^\mu \theta   \right)=0\,.
\end{align}
The asymptotic analysis shows that the non-normalizable and the normalizable
modes of the gauge field behave as 
\begin{equation}\label{behav}
A_{i(N.N.)}\sim A_{i(0)} r^{\Delta}\,;\hspace{1.5cm}A_{i(N.)}\sim\tilde A_{i(0)}
r^{-2-\Delta}\,; \hspace{1.5cm} \Delta=-1+\sqrt[]{1+m^2}\,.
\end{equation}
Since the mass has to be positive (for the massless case case saturates the
unitarity bound), there is no possible alternative quantization and the leading
term is always to be identified with the non-normalizable (N.N.) mode. Moreover,
there is an upper bound to the value of the mass prescribed by $\Delta=1$. As we
will show via holographic renormalization, the operator dual to the coefficient
of the non-normalizable mode
  is essentially given by the normalizable mode. Its dimension can be found via
the following argument. The AdS metric is invariant 
under the scaling $r\rightarrow \lambda r$, $(t,\vec{x}) \rightarrow
\lambda^{-1} (t,\vec{x})$. Since a gauge field is a one form
we have to study the behavior of $A_\mu(r,x) dx^\mu$ under these scalings one
finds then that the normalizable mode has a scaling
dimension of 
\begin{equation}
 \label{eq:dimoperator}
 \mathrm{dim}(\tilde A_{i(0)} ) = [J_i] = 3+\Delta\,.
\end{equation}
 This implies that if $\Delta >1$ the dual operator is irrelevant (in the IR)
and thus destroys the Ads asymptotics. In the holographic renormalization in
appendix \ref{app:ren} we find that accordingly the number of counterterms
diverges for $\Delta\rightarrow1$\footnote{We thank Ioannis Papadimitriou for
pointing this out.}. \\
It is clear that the number of counterterms depends on the value of the mass.
From now on we will work in the range of masses that minimizes it, namely
\begin{equation}
\Delta<\frac{1}{3} \longleftrightarrow m^2<\frac{7}{9}.
\end{equation} 
Henceforth we will refer to $\Delta$ as the anomalous dimension of the dual
current. \\
The procedure of renormalization for this theory is explained and discussed in
detail in the appendix \ref{app:ren}. The boundary action with the counter-terms
such that $S_{\mathcal{R}en}= S+S_{CT}$ reads
\begin{equation}
S_{CT}=\int_\partial d^4x\,\, \sqrt[]{-\gamma}\left( \frac{\Delta}{2}B_{i}B^{i}
- \frac{1}{4(\Delta+2)}\partial_i B^i \partial_j B^j  +
\frac{1}{8\Delta}F_{ij}F^{ij}       \right)\,,
\end{equation} 
with $B_i\equiv A_i-\partial \theta$.\\

Remarkably, the coupling of the St\"uckelberg field to the Chern-Simons  term in
(\ref{eq:1}) is not optional once the mass is turned on; if one does not add it
to the action it appears as a counterterm when holographic renormalization is
carried out.

\subsection{The one-point function}
From the renormalized action we compute correlators of the dual operators in the
boundary theory by means of the usual prescription. In this section we show our
results, sticking only to the strictly necessary technical details for the
discussion. A thorough version of the calculations can be found in appendix
\ref{app:1pfu1}. \\

Due to the anomalous dimension of the operator the analysis of the 1-point
function becomes more subtle than in the massless case. In previous works, at
zero mass, the leading terms of the expressions where finite. Therefore it made
sense to look at expression for the current VEV as a functional of the covariant
fields before taking the limit $r\rightarrow \infty$. This is however not the
case when $m\neq0$, since now all terms are divergent to leading order.
Nevertheless, to make comparison with the results at zero mass, we want to look
at the result before explicitly taking the limit. In order to do so, we split
the unrenormalized 1-pt. function into a term lacking a (sub leading) finite
contribution (called $X$ below) and terms which do lead to a finite contribution
after renormalization
\begin{align}\label{currentx}
\langle J^i \rangle= \lim_{r\rightarrow \infty} \,\,\sqrt[]{-g} r^\Delta  \left(
F^{ir} + r \Delta A^i \right) + X^i\,.
\end{align}
Remarkably we see that the contribution arising from the Chern-Simons term in
(\ref{eq:1}) is contained in $X^i$, which means that it does not contribute
\textbf{explicitly} to the current. The renormalized one-point function reads
\begin{align}
\label{tildea}
\langle J^i \rangle_{ren.}= 2(1+\Delta)\tilde A^i_{(0)}\,,
\end{align}
where $\tilde A_{(0)m}$ is the coefficient of the normalizable mode. Let us
compare this with the expression for the \textbf{consistent} current that one
obtains in absence of mass\footnote{Notice that in the zero mass limit $\theta$
becomes a non-dynamical
field defined at the boundary. The divergence of this field also contributes to
the current \cite{Landsteiner:2012kd}. In order to keep the discussion simple we
chose to take this non-dynamical field to vanish since this is the natural value
that arises from our background in the the zero mass limit.}
\begin{align}\label{currentm0}
\nonumber \langle J^i\rangle&= \lim_{r\rightarrow \infty} \,\,\sqrt[]{-g}  
\left( F^{ir} +\frac{4\kappa}{3} \epsilon^{ijkl}  A_j F_{kl} \right)+X^i\\
&\overset{\text{Ren.}} = 2 \tilde A^i_{(0)} +\frac{8\kappa}{3} \epsilon^{ijkl} 
A_{j(0)} \partial_k A_{l(0)}.
\end{align}
Here we see that in the massless case ($\Delta= 0$) the Chern Simons term indeed
gives a finite contribution to the current which is explicitly proportional to
the sources. It is precisely this term what makes the difference between the
covariant and the consistent definition of the current. We remind the reader
that in the case of global anomalies one can define a covariant current by
demanding that it transforms covariantly
under the anomalous gauge transformation. In the AdS/CFT dictionary this
covariant current is given by the normalizable mode of the vector
field. In contrast the consistent current is defined as the functional derivation
of the effective action with respect to the gauge field and
in the AdS/CFT correspondence includes the Chern-Simons term in
(\ref{currentm0}).

Equations (\ref{currentx}) and (\ref{tildea}) establish that we are no longer
able to make such a distinction if $m\ne 0$, for there is no explicit finite
local contribution of the Chern Simons term to the current operator. 
Quite remarkably, all of our results show that (\ref{tildea}) corresponds to
the \emph{consistent} current in the zero mass limit. This ultimately implies
that in the limit $m\rightarrow0$ the highly non-local expression $\tilde
A_{(0)i}$ gives rise to the two terms in the last line of (\ref{currentm0}),
which include a local term in the external sources. Hence, along the paper we
will only refer to consistent or covariant currents when analyzing the massless
limit.  
 
Another remarkable difference with the massless model is the Ward identity of
the current operator. Using the equations of motion we can write the divergence
of the current on-shell
\begin{align}
\label{dcurrent}
\nonumber \langle \partial_i J^i \rangle&= \lim_{r\rightarrow \infty}
\,\,\sqrt[]{-g} r^\Delta \left( m^2 \partial^r \theta + r\Delta  \partial_i  A^i
-\frac{\kappa}{3}\epsilon^{ijkl}F_{ij}F_{kl}+ \tilde
X\right)\\&\overset{\text{Ren.}}= 2(1+\Delta) \partial_i \tilde A^i_{(0)}
\,,\end{align}
where we have extracted the (infinite) Chern Simons term from
(\ref{currentx})\footnote{In other words, $\partial_i X^i =
-\frac{\kappa}{3}\epsilon^{ijkl}F_{ij}F_{kl}+ \tilde X$.} because it is
convenient for the following discussion. 
As mentioned before, the fact that the terms in these expressions diverge
obscures the interpretation if one does not take the limit $r\rightarrow
\infty$. Once we take it we find that the ward identity (\ref{dcurrent}) becomes
a tautology since the only term on the
right hand side that give a finite contribution is determined in the large $r$
expansion directly by the divergence of the normalizable 
mode of the vector field. 
Therefore the
divergence of the current on-shell is unconstrained. \\ If we now look at what
happens when we take the limit $m\rightarrow0$ before we take $r\rightarrow
\infty$ we see that we recover the expression for the divergence of the
consistent current
\begin{align}
\langle \partial_i J^i \rangle&= \lim_{r\rightarrow \infty} \,\,\sqrt[]{-g}
\left(\frac{\kappa}{3}\epsilon^{ijkl}F_{ij}F_{kl}+ \tilde
X\right)\overset{\text{Ren.}} =\sqrt{-g}
\frac{\kappa}{3}\epsilon^{ijkl}F_{ij}F_{kl}  \,.
\end{align}
%with the only finite contribution coming form the C.S. term. \\
%This implies that in the massive case the contribution of the C.S. term is
contained in the (non-local) normalizable mode of the filed. As we will see now
the behavior of the conductivity points in the same direction. 
\subsection{Two-point functions \& anomalous conductivity}
\label{subsec:axial2pt}
Our main interest is to study the effect that the anomalous dimension has on the
response of the system in presence of a magnetic field. As a first step in this
direction we compute the anomalous conductivity $J_i=\sigma_{55} B_i$ that is
related to a correlator of current operators via the Kubo formula
\begin{equation}\label{eq:s55}
\sigma_{55}=\left. \lim_{k\rightarrow 0}\frac{i}{k_z}\langle J_x J_y 
\rangle\right|_{\omega=0}.
\end{equation}
We emphasize however, that $B_i$ does not have the simple interpretation of a
magnetic field since its dimension is $2-\Delta$.
We want to study the anomalous conductivity in an analogous fashion to
\cite{Gynther:2010ed} and find the dependence of the chiral anomalous
conductivity on the source for $J^{\mu}$. 
In order to generalize the concept of chemical potential to the situation at
hand we
 we switch on a temporal component of the gauge field in the background 
$A=\Phi(r)\,dt$. We choose the axial gauge $A_r=0$. The equation of
motion is 
\begin{align}
\label{eq:back}
\Phi'' + \frac{3}{r}\Phi' - \frac{m^2}{f}\Phi=0\,.
\end{align}
We solve the equation (\ref{eq:back}) numerically\footnote{The analytic solution can be
worked out in terms of hypergeometric functions. Since we need to resort to numerical
methods lateron, when studying fluctuations around the background we found it more convenient to 
apply purely numerical methods also for the background.}, with the following boundary
conditions
\begin{align}\label{bconditions}
\phi(r_H)=0\,; \hspace{2cm} \phi(r\rightarrow\infty) \sim \mu_5 r^\Delta\,,
\end{align} 
with $\mu_5$ being the source. Notice that $\mu_5$ does not correspond to a
thermodynamic parameter in our massive model. Rather it 
should be interpreted as a coupling in the Hamiltonian. Different values of
$\mu_5$ correspond therefore to different theories.
Different values of a chemical potential correspond only to different filling
levels of the low lying
fermionic states in the same theory. In the case of an anomalous symmetry one
has to distinguish this filling level from the
presence of a background constant temporal component of the gauge field
\cite{Gynther:2010ed}.  

 The near horizon analysis shows that we are forced to impose $\Phi(r_H)=0$. In
absence of the mass term the gauge filed is not divergent at the horizon,
independently of the finite value it takes at the boundary. This reflects the
remanent (recall we work in the axial gauge) gauge freedom that one has in this
case: the value of the source can be shifted by a gauge
transformation\footnote{{Gauge transformations that are non-zero at the horizon
are not true gauge transformations but global transformations!}}. However the
mass term in the e.o.m. is divergent at the horizon and forces the field to
vanish there. 
Remarkably, this and the asymptotic behavior of the field illustrate the fact
that speaking of a chemical potential does no longer make sense. Computing the
chemical potential as the integrated radial electric flux in the bulk one
obtains
\begin{equation}
\mu= \lim_{r\rightarrow\infty}\int_{r_H}^r \partial_r A_t dr \rightarrow \infty.
\end{equation}
This can be understood heuristically from the non-conservation of the charge:
the energy to introduce and maintain a quantum of charge that is not conserved is
infinite. \\
Since our background is homogeneous in the transverse directions it is easy to
see that $\langle \partial_i J^i\rangle=0$. In particular, the fact that a
stationary solution exists implies that it is possible to choose a homogeneous
configuration of $\mu_5$ such that it compensates for the decay of the charge
that is naturally caused due to the mass term. Namely, 
\begin{align}
\label{eq:chdens}
\frac{d \rho}{dt}=0\,,
\end{align}
being $\rho$ the charge density of the system. Interestingly, we will see that
the source necessary to ensure (\ref{eq:chdens}) equals the axial chemical
potential in the massless limit (recall that only when $m=0$ we can identify
$\mu_5$ with a chemical potential). \\

Once we have built the background we can proceed to switch on perturbations on
top of it in order to compute the 2-point function (\ref{eq:s55}). To linear
order in the external source $\tilde{A}^i_{(0)} \approx\tilde{A}^i_{(0)} +
\delta \tilde a^i_{(0)}$. From (\ref{tildea}) we have 
\begin{align}\label{mirala}
\langle J^n J^m \rangle= 2(1+\Delta)\eta^{ml}\frac{\delta \tilde
a_{(0)l}}{\delta a_{(0)n}}\,.
\end{align}
%%%%%%%%%%%%%%%%%%%%%%%%%%%%%%%%%%%%%%%%%%%%%%
\begin{figure}[t!] 
\centering
\includegraphics[width=210pt]{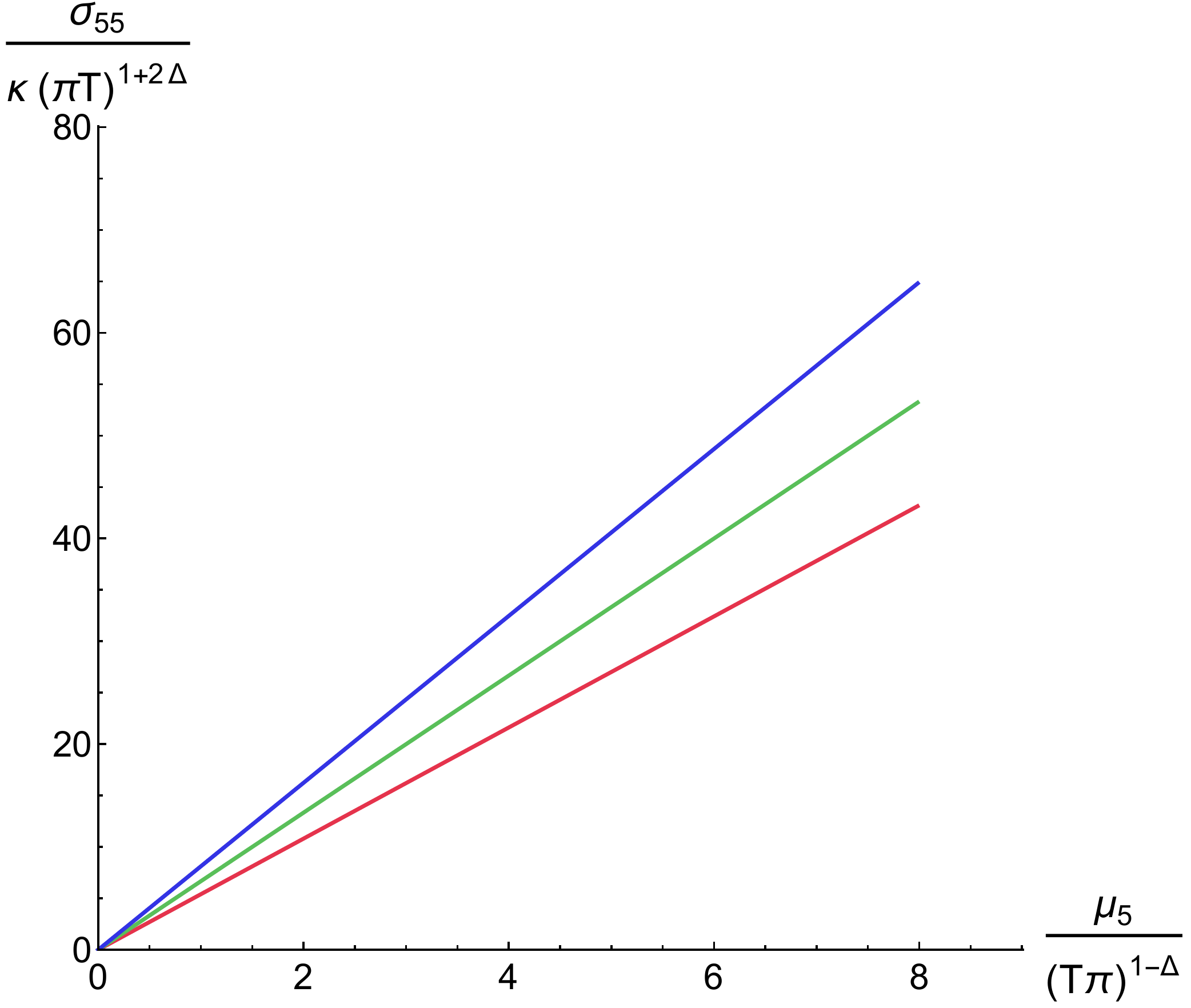}
\hspace{1cm}
\includegraphics[width=210pt]{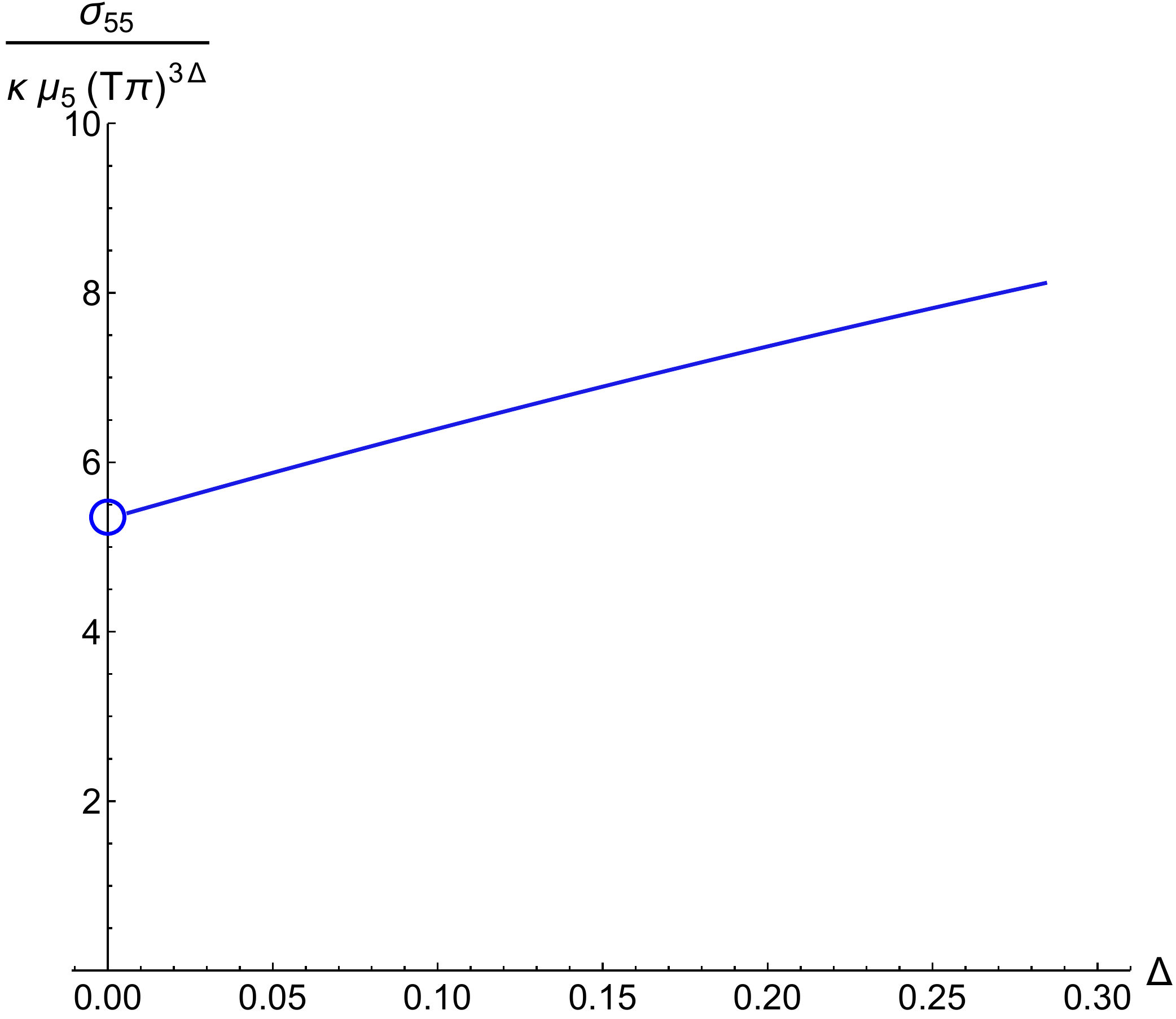}
\caption{\label{fig:s55u1}Left: Plot of the conductivity versus the source for:
$m^2=1/2$ (Blue), $m^2=1/4$ (Green), $m^2=0.01$ (Red). Right: Plot of the
conductivity coefficient as a function of the anomalous dimension; the circle
stands for the asymptotic value in the limit $\Delta\rightarrow 0$. }
\end{figure}
%%%%%%%%%%%%%%%%%%%%%%%%%%%%%%%%%%%%%%%%%%%%%%
%Where $ \tilde a_{(0)m}$ is the coefficient of the normalizable mode of the
perturbation.
We compute the above expression numerically. Again we leave technical details
for appendix (\ref{app:2pfu1}) because the analysis is tedious and is based on
standard techniques. We show the result in figure \ref{fig:s55u1}. A comment is
in order here regarding the temperature dependence on the plots. Dimensional
analysis of the correlator $[\langle J J\rangle]=6+2\Delta$ implies that the
conductivity has now dimension $[\sigma]=1+2\Delta$. This in turn causes the
physical conductivity to have a temperature dependence $\sigma \sim
T^{3\Delta}$. As usual, from numerics we can only plot dimensionless quantities
$\sigma/(\pi T)^{3\Delta}$ and $\mu_5/(\pi T)^{1-\Delta}$. 

 The plot on the left panel of figure \ref{fig:s55u1} shows the dependence of
the conductivity with the source for different values of the mass. Despite the
fact that the slope changes the behavior is always linear on the dimensionless
source. The plot on the right panel shows the conductivity coefficient vs. the
anomalous dimension of the current $\Delta$. Remarkably the conductivity gets
enhanced by the presence of the mass term in the bulk. In addition to this
enhancement the plot shows another feature that deserves a comment. In the limit
$\Delta(m)\rightarrow0$ the conductivity goes to the numerical value $5.333
\sim \frac{16}{3}$. Let us now look at the analytic solution for zero mass shown
in (2.25) of \cite{Gynther:2010ed}\footnote{This model contained two massless
vector fields in the bulk. One modeling
the conserved vector and the other the anomalous axial symmetry. It is clear
that the result obtained in our model with one massive vector
should be compared in the zero mass limit to the axial vector sector of the
model in \cite{Gynther:2010ed}.}
\begin{equation}\label{eq:oldresult}
\langle J_5^i J_5^j\rangle=-4 i \tilde\kappa k (3\mu_5-\alpha)\epsilon_{ij}\,,
\end{equation}
where $\mu_5$ here is the thermodynamic chemical potential, $\alpha$ is the
source, i.e. the boundary value of the temporal component
of the gauge field and $\tilde \kappa = \frac{2\kappa}{3}$ in our convention. If
one chooses the gauge $\alpha=\mu_5$ then one obtains
\begin{equation}
\langle J_5^i J_5^j\rangle=-8 i \tilde\kappa k \mu_5\epsilon_{ij}\,,
\end{equation}
\begin{figure}[t!] 
\centering
\includegraphics[width=250pt]{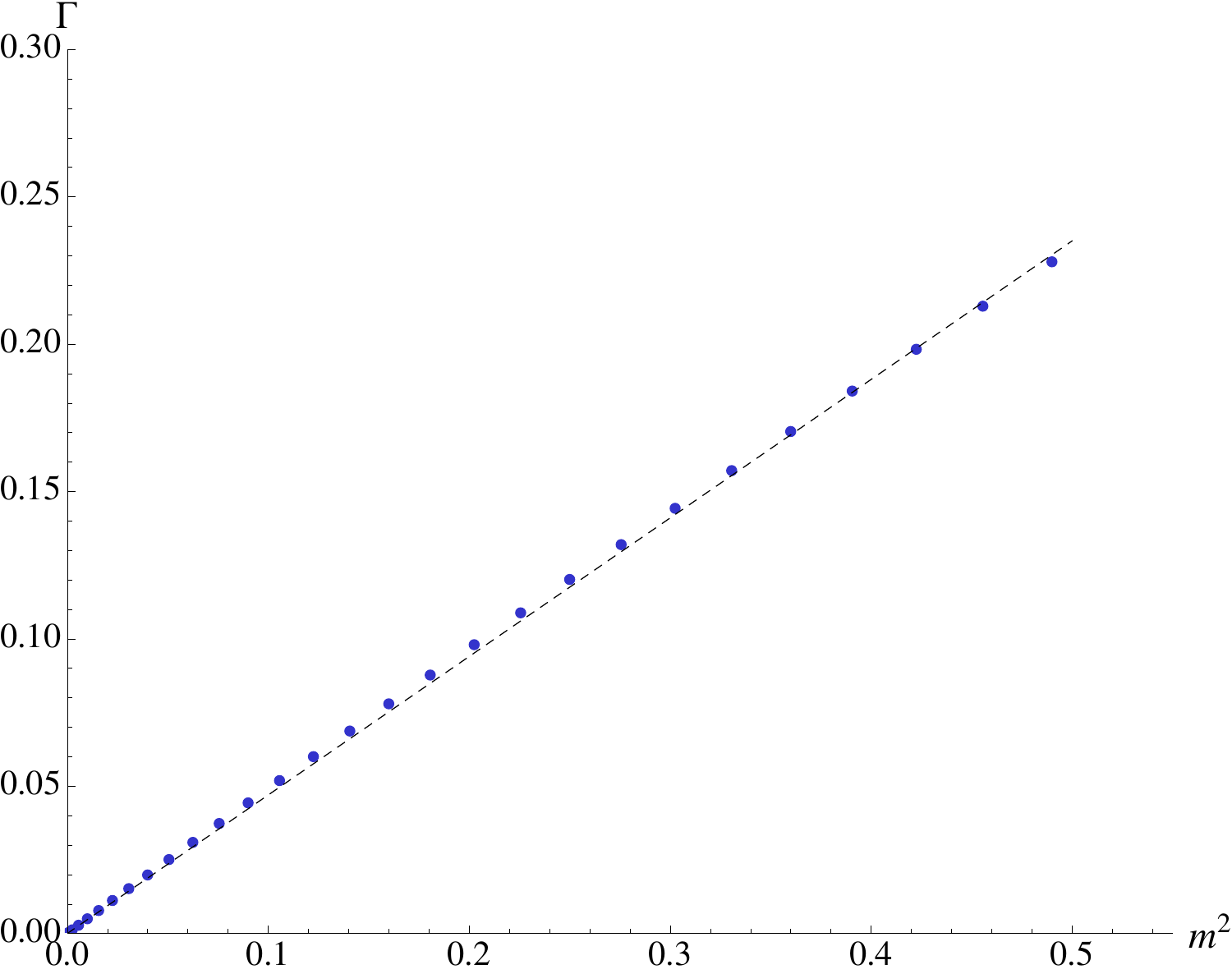}\hspace{2mm}
\caption{\label{fig:gapu1} The gap $\Gamma$ versus $m^2$. Black line corresponds
to a linear fit.}
\end{figure}
In our numerical results we have absorbed the Chern Simons coupling in the
definition of the external B-field (or equivalently set
it to one in the fluctuation equations). Taking into account the difference in
the normalizations of the Chern Simons couplings
in \cite{Gynther:2010ed}  we can extract from (\ref{eq:oldresult}) the numerical
value $\sigma_{55} = 16/3$ which coincides with the $m\rightarrow 0$ limit in
our case!
 We conclude that our result matches the analytic formula only if we identify
$\alpha=\mu_5$, as mentioned right after equation (\ref{eq:chdens}). This is
also consistent with what we found in the expression for the current. 

The fact that we found a time-independent background solution obscures
the non-conservation of the charge. The best way to shed light onto the explicit
decay of charge is by considering a trivial background (in particular, all the
sources vanish) and look at the spectrum of quasinormal modes. 
In the massless case the lowest QNM shows a diffusion-type behavior, namely
$\omega= -iDk^2$. This diffusive mode has to develop a gap when $m\neq0$ due to
the non-conservation of the charge. Technical details on how to compute QNM can
be found in \cite{Kaminski:2009dh}. 
Indeed we find that the lowest QNM is no longer massless. The gap $\Gamma$
depends on the value of the bulk mass as depicted in figure \ref{fig:gapu1}.

This indicates that the charge is no longer conserved. Furthermore a simple
phenomenological model including only the dynamics of the
lowest quasinormal mode suggests that the non-conservation can be modeled by
writing $\partial_\mu J^\mu = -\frac 1 \tau J^0$, where $\tau$ is
the gap of the lowest quasinormal mode. Indeed, such a phenomenological decay law
together with Fick's law $\vec{J} = - D \vec\nabla J^0$ suggests
a gapped pseudo diffusive mode $\omega + i/\tau + i D k^2 =0$ which indeed is
what we find from the QNM spectrum (see next section).

%%%%%%%%%%%%%%             The U(1)xU(1) Model             %%%%%%%%%%%%%%%%%%

\section{The St\"uckelberg U(1)xU(1) model}
\label{sec:U1U1model}
In this section we introduce an extra unbroken abelian symmetry in the bulk.
This allows us to switch on an ``honest'' external magnetic field in the dual
theory and therefore study not only the axial conductivity but the Chiral
Magnetic conductivity and the Chiral Separation conductivity as well. In addition
we will be able to study the effect the mass has on the Chiral Magnetic Wave and
on the electric conductivity. The Lagrangian reads
\begin{align}
\label{actionu1u1}
\mathcal{L}=\left(
-\frac{1}{4}F^2-\frac{1}{4}H^2-\frac{m^2}{2}(A_\mu-\partial_\mu
\theta)(A^\mu-\partial^\mu
\theta)+\frac{\kappa}{2}\epsilon^{\mu\alpha\beta\gamma\delta}(A_\mu-\partial_\mu
\theta)\left(F_{\alpha\beta}F_{\gamma\delta}+3 H_{\alpha\beta}H_{\gamma\delta} 
\right)  \right)
%S&=\int\sqrt[]{-g}\left(
%-\frac{1}{4}F^2-\frac{1}{4}H^2-\frac{m^2}{2}(A-d\theta)\wedge *
%(A-d\theta)+\frac{\kappa}{2}(A-d\theta)\wedge\left( F\wedge F+3 H\wedge H 
%\right)  \right)\,,
\end{align}
where $F=dA$ and $H=dV$. The new dynamical U(1) in the bulk is massless and
couples to the Chern-Simons term in the usual way. As in the previous section we
work in the probe limit with Schwarzschild-$AdS_5$ as background metric. The
scalar field transforms non-trivially only under the massive $U(1)$.  From now
on we will refer to the massless $U(1)$ as ``vector'' $V_\mu$ and to the massive
$U(1)$ as ``axial'' $A_\mu$.
% in reference to the (non-)conservation of the consistent currents associated
%to them when the mass is zero
%\begin{align}
%\langle \partial_i J_V^i\rangle=0\,,\hspace{2cm}
%\langle \partial_i J^i_A \rangle=\kappa \left(F\wedge F + 3 H\wedge H 
%\right)\,.
%\end{align}
 The equations of motion for the gauge fields are
\begin{align}
\nabla_\mu F^{\mu \nu} - m^2 (A^\nu-\partial^\nu\theta ) +
\frac{3\kappa}{2}\epsilon^{\nu \alpha \beta \gamma \rho} (F_{\alpha \beta}
F_{\gamma \rho}+H_{\alpha \beta} H_{\gamma \rho}) =0\,, \\
\nabla_\nu H^{\nu \mu}+ 3 \kappa \epsilon^{\mu \alpha \beta \gamma \rho}
F_{\alpha \beta} H_{\gamma \rho}  =0\,.
\end{align}
The equation of motion of the scalar remains unchanged (see equation
(\ref{eq:scalareq})). Non-normalizable and normalizable modes of the axial gauge
field have the same asymptotics for large $r$ as the gauge field in the $U(1)$
model. The vector field shows the same behavior at infinity as usual
\begin{equation}\label{behavu2}
V_{i(N.N.)}\sim A_{i(0)} r^{0}\,;\hspace{1.5cm}V_{i(N.)}\sim\tilde V_{i(0)}
r^{-2}\,.
\end{equation}
The holographic renormalization of this model is discussed in appendix
\ref{app:ren2}. The result is the following boundary term
\begin{equation}
S_{CT}=\int_\partial d^4x\,\, \sqrt[]{-\gamma}\left( \frac{\Delta}{2}B_{i}B^{i}
- \frac{1}{4(\Delta+2)}\partial_i B^i \partial_j B^j  +
\frac{1}{8\Delta}F_{ij}F^{ij} +\frac{1}{8}H_{ij}H^{ij}  \log r^2    \right)\,,
\end{equation} 
with $B_i = A_i -\partial_i \theta$. There are two differences from the result
in the previous model. On the one hand, the appearance of the usual $\sim \log$
term for the vector gauge field. On the other, the role of the coupling of the
St\"uckelberg field to the C.S term in (\ref{actionu1u1}) is different because
now we have two independent couplings $d\theta\wedge F\wedge F$ and
$d\theta\wedge H\wedge H$. The former is mandatory, as in the the $U(1)$ model.
The latter however is optional since it is a finite boundary term\footnote{At
zero mass this coupling corresponds to the axion term discussed in
\cite{Landsteiner:2011tf}.}. We have chosen to include it. As we will see, this
will not affect the results in our concrete background, but it is potentially
useful for other models since it cancels possible finite contributions to the
vector current stemming from the St\"uckelberg field.

%%%%%%%%%%                      CME & CSE                   %%%%%%%%%%%%%

\subsection{One-point functions}
\label{subsec:1ptU1U1}
First we compute the 1-point functions of the gauge fields. The technical
details of the calculations can be found in appendix \ref{app:1pfu1u1}. As in
(\ref{currentx}) we hide all terms that do not contain any finite contribution
in vectors $X^i$ and $Y^i$, obtaining
\begin{align}\label{correintev}
\langle J_V^i \rangle&= \lim_{r\rightarrow \infty} \,\,\sqrt[]{-g}
\left(H^{ir}+6\kappa \epsilon^{ijkl} \left( A_j-\partial_j \theta \right)H_{kl}
\right) +X^i\,,  \\ \label{correintea}
\langle J^i_A \rangle&= \lim_{r\rightarrow \infty} \,\,\sqrt[]{-g} r^\Delta 
\left( F^{ir} + r \Delta A^i \right) + Y^i\,.
\end{align} 
The axial current behaves as in the previous model. 
%The vector current looks precisely like the one found at $m=0$
%\begin{align}
%\langle J_V^i \rangle&= \lim_{r\rightarrow \infty} \,\,\sqrt[]{-g}
%\left(H^{ir}+6\kappa \epsilon^{ijkl}  A_jH_{kl} \right) +X^i\,.
%\end{align} 
Recall that the leading term in the asymptotic expansion of the axial gauge
field diverges and so it does the Chern Simons term in (\ref{correintev}).
Nevertheless, contrary to the axial current, this term has a subleading finite
part (which is the reason why we do not
include it in $X^i$). Looking at the complete expansion for the axial gauge
field (\ref{expansion}), we see that this finite contribution is proportional to
the source of $\theta$ instead of the source of the gauge field. This is of
course different from what one finds in the massless case. In addition, it is
here where we see the effect that the coupling $d\theta\wedge H\wedge H$ has. It
cancels this finite contribution proportional to the source dual to the
St\"uckelberg field. As mentioned before, this cancellation comes from the
choice we made in the action and can be removed at will.\\ 
We can now look at the Ward identities. Substituting the e.o.m. in the
divergence of the current we find
\begin{align}
\langle \partial_i J^i_V \rangle_{\text{Ren.}}=0\,; \hspace{2cm}    \langle
\partial_i J^i_A \rangle_{\text{Ren.}}= (2+2\Delta) \partial_i \tilde
A_{i(0)}\,.
\end{align}
The vector current is conserved as in the massless case. The result for the
axial current is the same as in the previous model: its divergence is
unconstrained reflecting the fact that it is a non-conserved current.

\subsection{Two-point functions \& anomalous conductivities}
The presence of an extra $U(1)$ allows us to obtain the following anomalous
conductivities from Kubo formulae \cite{Gynther:2010ed,Amado:2011zx} 
\begin{align}
\sigma_{CME}=\lim_{k\rightarrow
0}\frac{i\epsilon_{ij}}{2k}\left<J^iJ^j\right>(\omega=0,k)\label{kubo}\,,\\
\sigma_{CSE}=\lim_{k\rightarrow
0}\frac{i\epsilon_{ij}}{2k}\left<J_5^iJ^j\right>(\omega=0,k)\label{kubos}\,,\\
\label{kuboss} \sigma_{55}=\lim_{k\rightarrow
0}\frac{i\epsilon_{ij}}{2k}\left<J_5^iJ_5^j\right>(\omega=0,k)\,. 
\end{align}
In order to study these we have to switch on a source for both axial and vector
charges. Since the vector charge is conserved at the boundary it is possible to
define a non-divergent chemical potential for it. In fact, since the vector
charge is conserved we do not need to source
it by a constant $V_0$ at the boundary. Formally $V_0$ is just a pure gauge and
therefore does not enter any physical observables. It is
however a convenient and standard choice to reflect the presence of the chemical
potential in the vector sector by choosing $V_0 = \mu$
at the boundary and $V_0 = 0$ at the horizon. In this case the difference of
potentials at the boundary and the Horizon is the energy
needed to introduce one unit of charge into the ensemble. This is a finite
quantity and by definition the chemical potential $\mu$.\\

We want to see how the dependence of the conductivities on the source and/or
chemical potential is affected by the mass. Our background consists of the
non-trivial temporal components of both gauge fields. It is static and
homogeneous in the dual theory so the bulk fields only depend on the radial
coordinate (again we work in the axial gauge $A_r=0,\ V_r=0$)
\begin{align}\label{bakbakbak}
\theta(r)=0;  \hspace{4mm} A = \phi(r) dt ;  \hspace{4mm} V = \chi(r)dt\,.
\end{align}
The equations to solve are
\begin{align}
\phi'' + \frac{3}{r}\phi' - \frac{m^2}{f}\phi=0\label{eq:phi}\,,\\
\chi'' + \frac{3}{r}\chi' =0 \label{eq:chi}\,.
\end{align} 
The boundary conditions for the gauge fields at infinity $\phi(r\rightarrow
\infty)=\mu_A r^{\Delta };\  \chi(r\rightarrow \infty)=\mu_V $ determine the
value of the sources. As usual, (\ref{eq:chi}) has the analytic solution
\begin{equation}
\chi(r)=\mu_V-\frac{\mu_V}{r^2}\,.
\end{equation}
Expanding the action to second order in the perturbations and differentiating
w.r.t. the sources we obtain the concrete expressions for the renormalized
correlators
\begin{align}
&\langle J_i^V J_j^V\rangle_{\text{Ren.}}= 2 \eta_{mj}\frac{\delta \tilde
v_{i(0)}}{\delta v_{m(0)}}\\
&\langle J_i^A J_j^A\rangle_{\text{Ren.}}= (2+2\Delta)\eta_{mj} \frac{\delta \tilde
a_{i(0)}}{\delta a_{m(0)}}\\
&\langle J_i^A J_j^V\rangle_{\text{Ren.}}= 2\eta_{mj} \frac{\delta \tilde
v_{i(0)}}{\delta a_{m(0)}} =  (2+2\Delta)\eta_{mj} \frac{\delta \tilde a_{i(0)}}{\delta
v_{m(0)}}
\end{align}
\begin{figure}[t!]
\centering
\includegraphics[width=210pt]{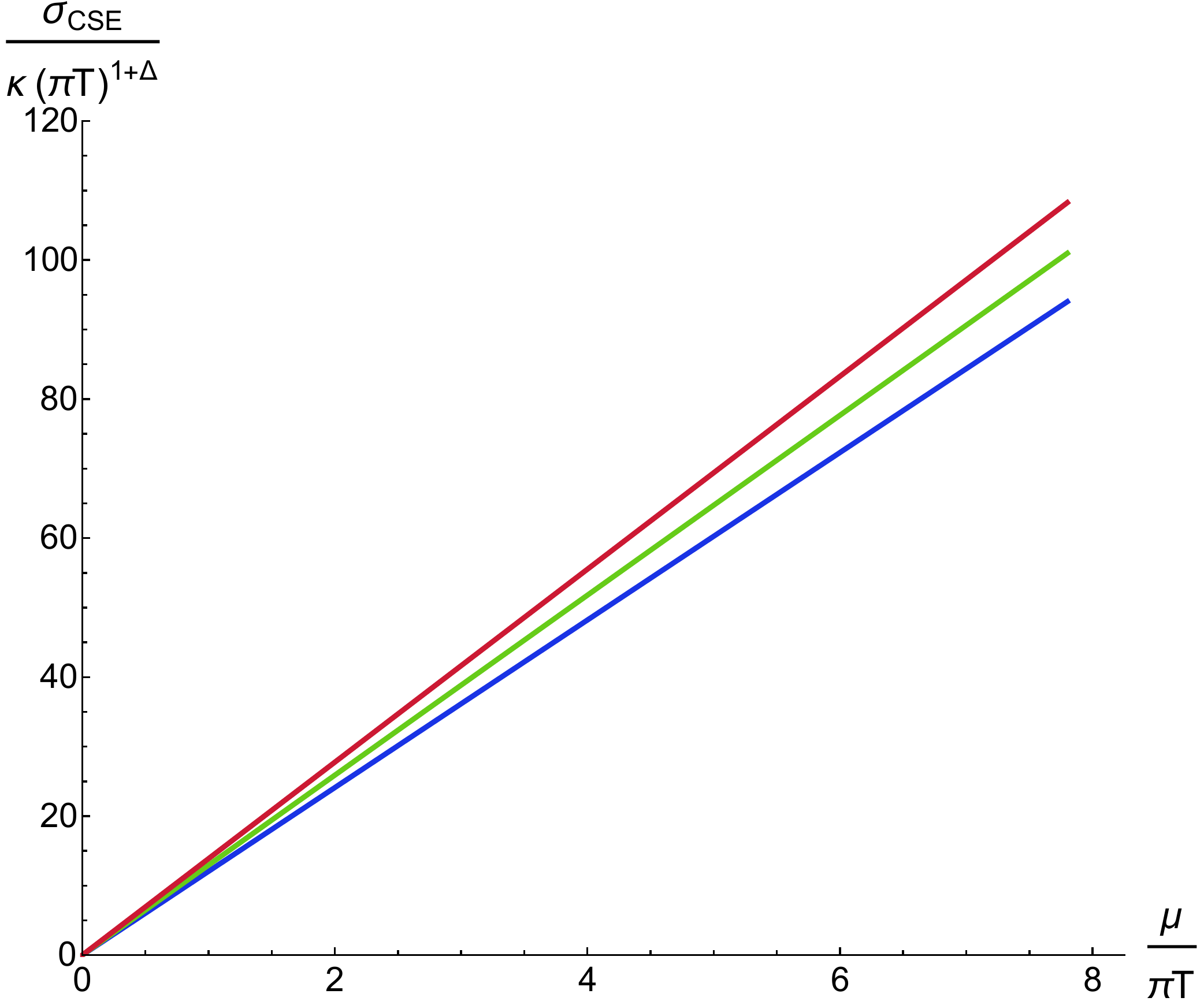}
\hspace{1cm}
\includegraphics[width=210pt]{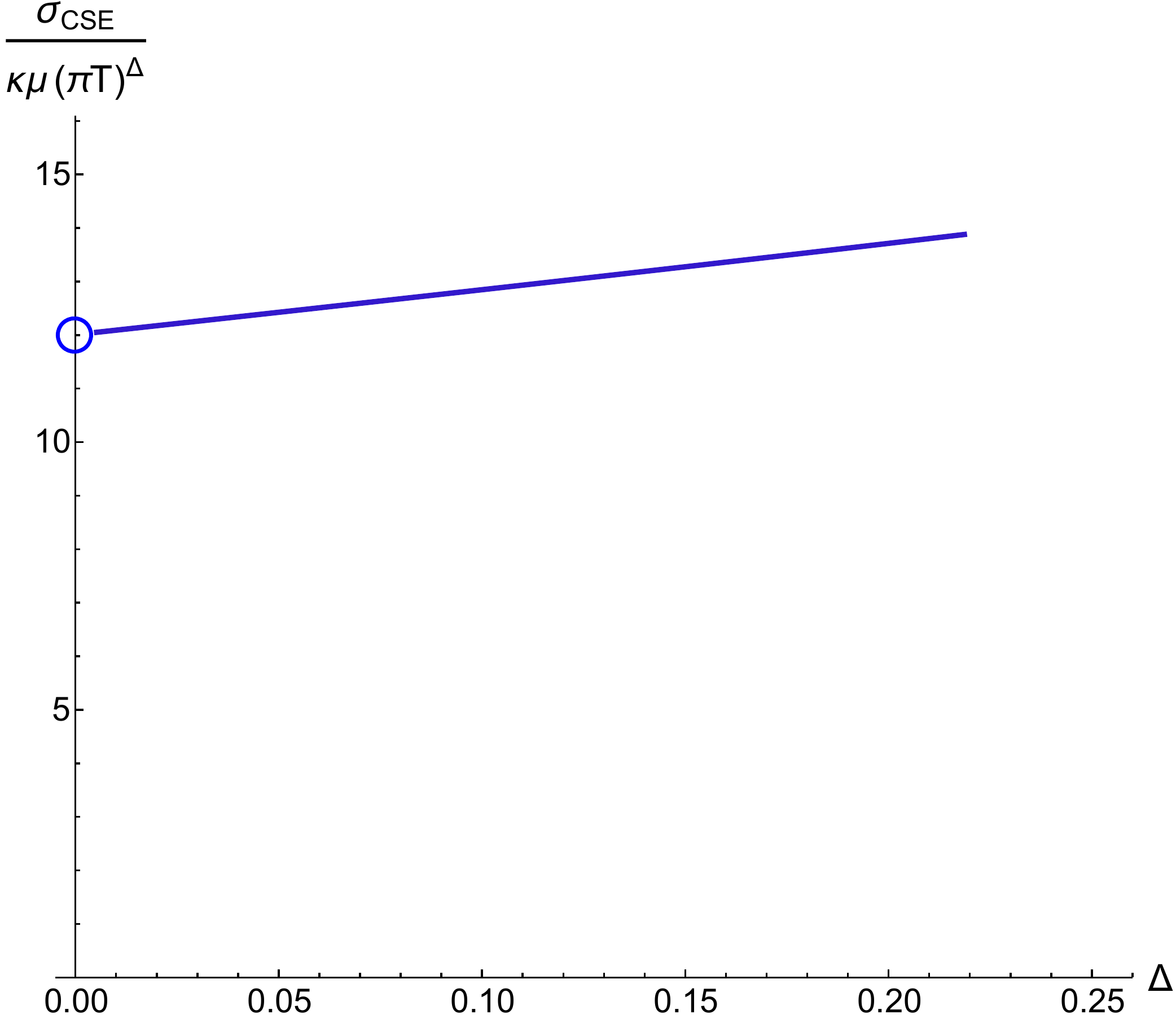}
\caption{ \label{fig:cse}Left: Plot of the CSC versus the chemical potential
$\mu$ for: $m^2=1/2$ (Blue), $m^2=1/3$ (Green), $m^2=0$ (Red). This conductivity
is independent of the axial source $\mu_5$. Right: Plot of the CSC coefficient
as a function of the anomalous dimension $\Delta= \sqrt[]{m^2+1}-1$.}
\end{figure}
We compute the above correlators numerically. For a detailed explanation see
appendix \ref{app:2pfu1u1}. In the following we comment on the outcome.
\paragraph{Axial Conductivity:} the conductivity $\sigma_{55}$ related to the
correlator of two axial currents behaves identically to section
\ref{subsec:axial2pt}. Hence, we refer the reader to figure \ref{fig:s55u1} and
the corresponding discussion.
\paragraph{Chiral Separation Conductivity:} we show the result in figure
\ref{fig:cse}. In the plot on the l.h.s. we show the behavior of the
conductivity with the vector chemical potential $\mu$. We find that there is no
dependence on the source $\mu_5$ for any value of the mass/anomalous dimension.
As in the axial conductivity we observe an enhancement with increasing mass. In
addition, in the massless limit the conductivity approaches the value
$\sigma_{CSE} \approx 12$ in numerical units. Again this is in agreement with
the analytic solution for $m=0$ \cite{Gynther:2010ed}. Notice that for this
conductivity even in $m=0$ there is no dependence on the value of the vector
field at the horizon (the source). 
\paragraph{Chiral Magnetic Conductivity:} the CME vanishes in our background.
This is in perfect agreement with all the findings so far. As it happened
with the rest of anomalous conductivities, in the massless limit the CMC
approaches the value that one obtains for the consistent currents. We believe
that the fact that it vanishes even in the massive case is a consequence of the
the presence of the source $\mu_5$.
The necessary source to achieve a stationary solution for any value of $m$ is
such that it forces the anomalous response of $J^i_V$ to $B^i_V$ to vanish, very
much as it occurs at zero mass. This does however not imply that the Chiral
Magnetic effect does not exist in this model.
As we will see in the following, if we allow the axial charge to fluctuate
freely (as opposed to fixing its value via a source term) the chiral
magnetic effect is realized. In particular it gives rise to a (generalization of
the) chiral magnetic wave and to a negative magneto resistivity.
Both of which can be understood as a manifestation of the chiral magnetic
effect.

%%%%%%%%%%%%%%%%%%%                CMW           %%%%%%%%%%%%%%%%%%%%%%%%%%%

\subsection{The Chiral Magnetic Wave}
\label{subsec:CMW}
We start be reviewing the essential features in the case when also the axial
current is a canonical dimension three current.
The chiral magnetic wave (CMW) is a collective massless excitation that arises
form the coupling of vector and axial density waves in presence of a magnetic
field \cite{Kharzeev:2010gd}. 
In addition, this mode can only appear in the spectrum if there is an underlying
axial anomaly. The dispersion relation for this mode corresponds to a damped
sound wave
\begin{align}
\label{dispersioncmw}
\omega(k)= \pm v_{\chi}k-i D k^2\,,
\end{align}
although it is related to transport of electric and axial charge. This mode can
be thought of as a combination of the CME and the CSE.
The vector charge and the axial charge oscillate one into the other giving rise
to a propagating wave. This wave mode is present 
 even in the absence of net axial or vector charge.  The CMW is expected to
play an important role in the experimental confirmations of anomaly induced
transport effects. It has been argued in the case of heavy ion collisions that
the CMW induces a quadrupole moment in
the electric charge distribution of the final state hadrons
\cite{Burnier:2011bf}. 

Let us analyze how this propagating mode is affected by the St\"uckelberg
mechanism in the bulk. Before we proceed to study holographic numerical results
we can perform a purely hydrodynamic computation as follows. 
As we have already shown in the previous section, the presence of the mass term
for the axial vector field leads to a non-vanishing, purely
imaginary gap for the lowest quasinormal mode. We will include this gap as a
decay constant for axial charge.
Consider thus a model with axial and vector symmetries. Under the assumption of
the existence of a $AVV$ anomaly in the system, the constitutive relations for
the current in the presence of a background magnetic field $B$ read
\begin{align}
j^x_V = \frac{\kappa\rho_A B}{\chi_A}- D \partial_x\rho_V\,; \hspace{3cm}
j^x_A = \frac{\kappa\rho_V B}{\chi_V} - D\partial_x \rho_A\,.
\end{align}
with $D$ the Diffusion constant and $\kappa$ the anomaly coefficient. We assume
CME and CSE to be present and have expressed them in
terms of the charge densities and the susceptibilities $\chi_A$, $\chi_V$
\cite{Kharzeev:2010gd}.
On the other hand we have the (non-)conservation equations
\begin{align}
\partial_{\mu} j_V^{\mu}=0\,;\hspace{4cm}
\label{eq:noncons}\partial_{\mu} j_A^{\mu}= -\Gamma\rho_A
\end{align}
Where $\Gamma(m)$ is the charge dissipation induced by the coupling to the
underlying gauge anomaly\footnote{We also assume
vanishing external electric field and therefore there is no $\vec{E}.\vec{B}$
term present in the equation for the axial current.}. 
From here we get the coupled equations
\begin{align}
\omega \rho_V + \frac{k\kappa\rho_A B}{\chi_A}+ik^2D\rho_V=0\,,\\
(\omega +i \Gamma) \rho_A + \frac{k\kappa\rho_V B}{\chi_V}+ik^2D\rho_A=0\,.
\end{align}
Assuming now that the equations are linearly dependent we get
\begin{align} \label{hydroluis}
\omega_{\pm}= -\frac{i\Gamma}{2} -iDk^2\pm\,
\sqrt[]{\frac{B^2k^2\kappa^2}{\chi_A \chi_V}    -\frac{\Gamma^2}{4}}\,.
\end{align} 
\begin{figure}[t!] 
\centering
\includegraphics[width=220pt]{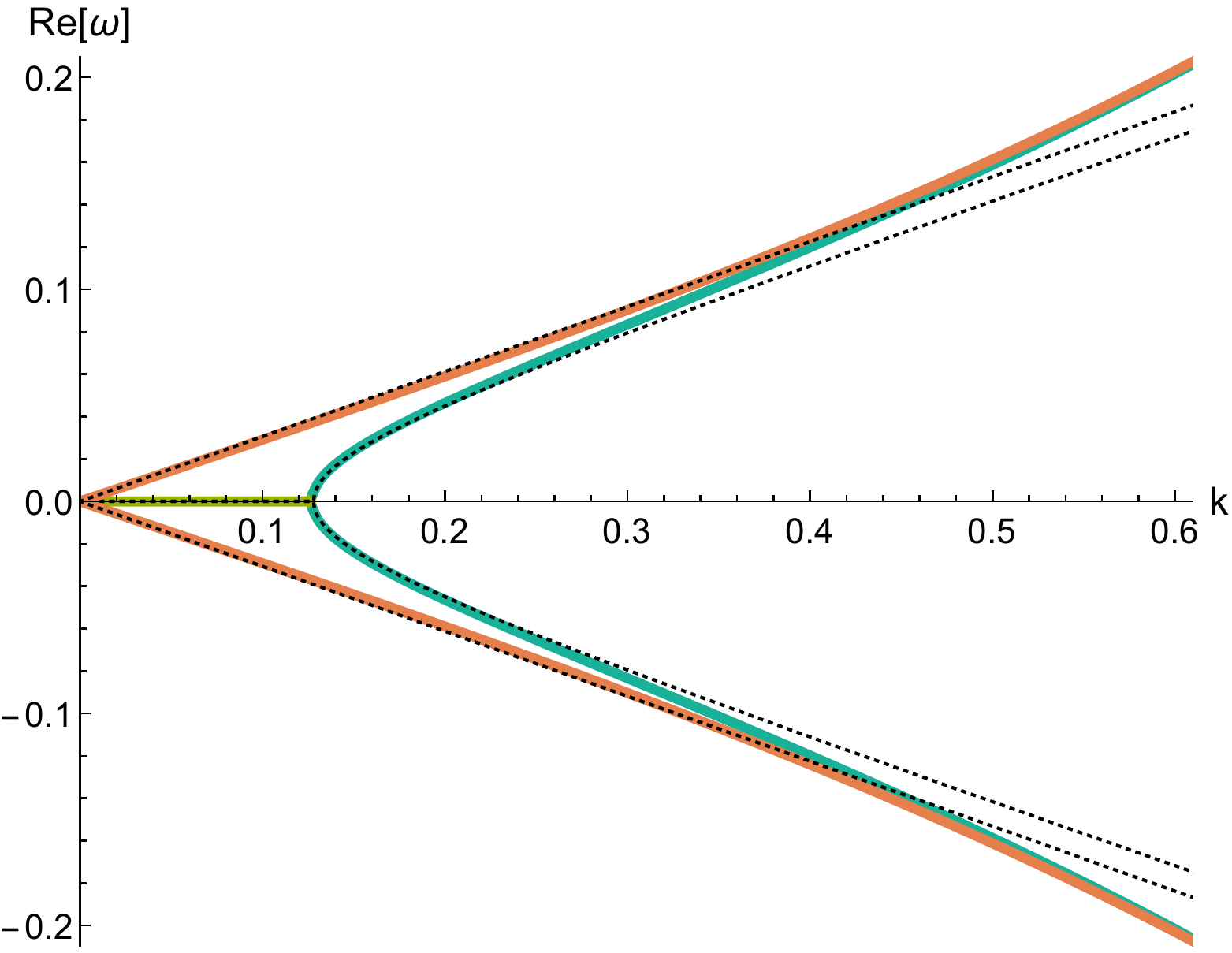}\hspace{2mm}
\includegraphics[width=220pt]{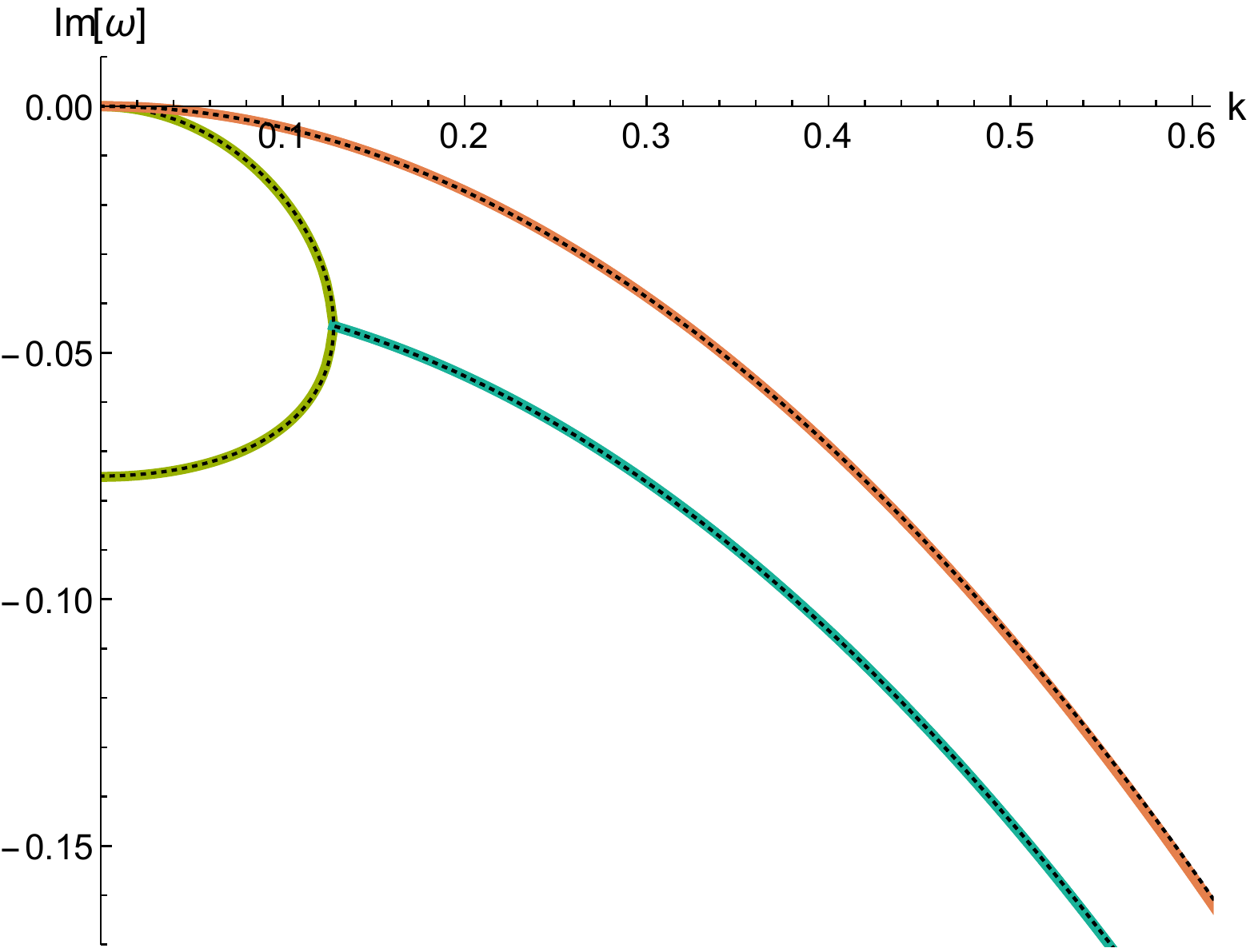}
\caption{\label{fig:REIM} (Color online) Real and Imaginary parts of the
frequency of the lowest QNM as a function of k. Solid lines correspond to
numerical data with $\kappa B=0.05$ and two different values of the mass:
$m^2=0$ (orange) and $m^2=0.15$ ($\Delta=0.08$) (blue, green). The Massive case
is given two different colors to highlight the regimes $k<k_c$ (green) and
$k>k_c$ (blue). Dashed lines correspond to the analytic formula
(\ref{hydroluis}). The massless case shows the behavior of the CMW. With a
non-vanishing mass such a behavior s recovered for $k>k_c$. }
\end{figure}
The mode associated to $\omega_+$ is massless and expected to arise due to the
fact that the vector symmetry is conserved. 
It basically represents the Diffusion law for the conserved vector charge.
The $\omega_-$ mode is gapped, i.e. $\omega_-(k=0)=-i\Gamma$. Both combine at a
critical value for the momentum $k_C(\Gamma, B, \chi_{(V,A)})$ given by the zero
of the term inside the square root.
\begin{itemize}
\item If $4B^2 \kappa^2 k^2 > \chi_A\chi_V \Gamma^2$ the square root is real and
we obtain a contribution linear in $k$ (which boils down to the well-known
linear dispersion relation of the Chiral Magnetic Wave in the limit
$\Gamma=0$). 
\item If $4B^2 \kappa^2 k^2 < \chi_A\chi_V \Gamma^2$ the square root
contribution is completely contained in the imaginary part of the frequency.
\end{itemize}
In summary, we see that 
\begin{align}
\label{eq:critk}
k_C =\frac{ \chi_A\chi_V \Gamma^2}{4B^2 \kappa^2}\,.
\end{align}
For $k>k_C$ we get a propagating mode whose dispersion relation approximates the one of the CMW\footnote{Observe that for $k>>k_c$ the slope
$\Re(\omega)/k$ is the same as in the case $\Gamma=0$.}. On the contrary, if
$k<k_C$, there is no real part of the frequency (i.e. no Chiral Magnetic Wave);
one of the modes remains massless and the other develops a gap $\Gamma$.\\
%%%%%%%%%%%%%%%%%%%%%%%%%%%%%%%%%%%%%%%%%%%%%%%%%%%%%%%%%%%%%%LUIS%%%%%%%%%%%%%%
%%%%%%%%%%%%%%%%

With this phenomenological model in mind we look for these modes in our
holographic model. In order to find the CMW we look at the QNM spectrum in
presence of a constant magnetic field $B$ in the $z$-direction. Since the CMW is
present at zero axial and vector charge densities, we do not switch on any
chemical potential in the background.
The only non-zero field in our ansatz for the background is $A_x=B y$.
It is easy to check that such an ansatz satisfies the equations of motion
trivially. Subsequently we study the perturbations, with momentum $k$ aligned
with the magnetic field.
\begin{figure}[h] 
\centering
\includegraphics[width=230pt]{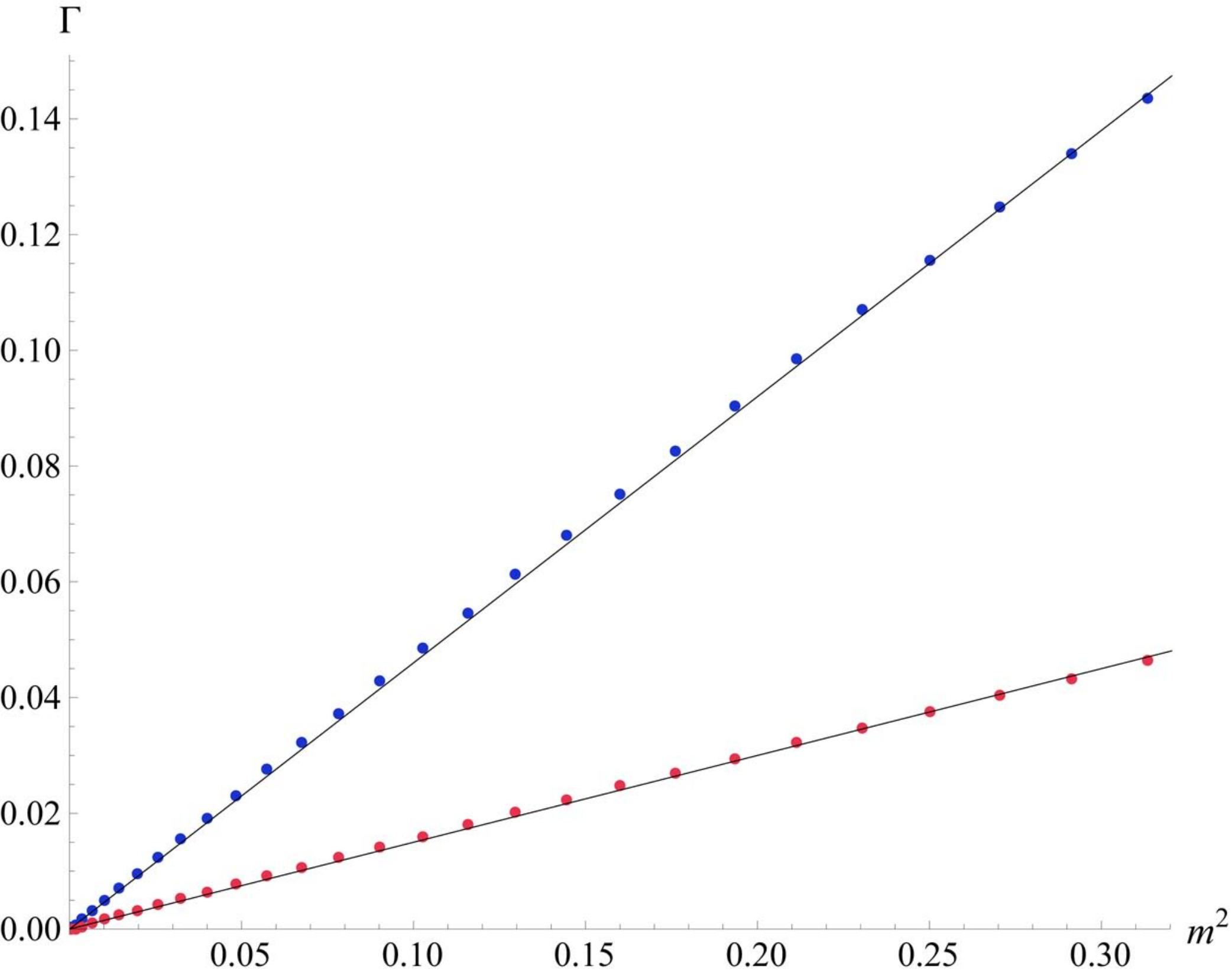}\hspace{2mm}
\caption{\label{fig:gama}(Color online) The gap $\Gamma$ versus $m^2$ for
different values of the magnetic field $\kappa B=0.01$ (blue) and  $\kappa
B=0.5$ (red) . Black lines correspond to linear fits.}
\end{figure}
Applying the determinant method of \cite{Kaminski:2009dh} we are able to obtain
the dispersion relation of the CMW as depicted in figure \ref{fig:REIM}; we show
the dispersion relation of the lowest QNMs for both $m=0$ (orange) and $m>0$
(green, blue) in presence of $B$. On top of this we plot (dashed lines) we show
a fit to the predictions of the phenomenological model (\ref{hydroluis}). \\
The numerical results agree very well with the analytic analysis and we observe
the appearance of a critical momentum $k_C$, induced by the mass term. Below
this momentum the Chiral Magnetic Wave is not really wave-like (i.e.
$\Re[\omega(k)]=0 \text{\ for\ } k<k_C $); the two modes decouple, giving rise
to a diffusive mode and gapped purely imaginary mode. Such a spectrum is what
one would expect to find in the model if there was no CMW, that is, the unbroken
vector charge exhibits diffusive behavior, with a massless mode protected by
the symmetry, whereas the analogous mode for broken axial U(1) symmetry develops
a gap $\Gamma$. This gap is proportional to the mass and gets diminished the stronger
the magnetic field becomes. Above the critical momentum the two modes fuse
again, giving rise to the expected behavior of the CMW.
Since the CMW is a propagating oscillation between axial and vector charge we
see that for small momentum the decay of the axial charge
dominates, i.e. the axial charge decays before it can oscillate back into vector
charge.The strength of the mixing of the charges
is proportional to the momentum. This mixing becomes large enough and the
oscillation fast enough to allow the build up of a propagating
(damped) wave at large enough momentum. 
\\

We show the behavior of the gap $\Gamma$ with the mass for different values of
the magnetic field in figure \ref{fig:gama}. We find that the gap goes as $\sim
m^2$ and that it is inversely proportional to the strength of the magnetic
field.
%%%%%%%%%%%%%%%%%%%%%%%%%%%            CMW &CONDUCTIVITIES        %%%%%%%%%%%%%%%

\subsection{Negative Magneto Resistivity}
\label{subsec:CMWeconds}
\begin{figure}[t!] 
\centering
\includegraphics[width=220pt]{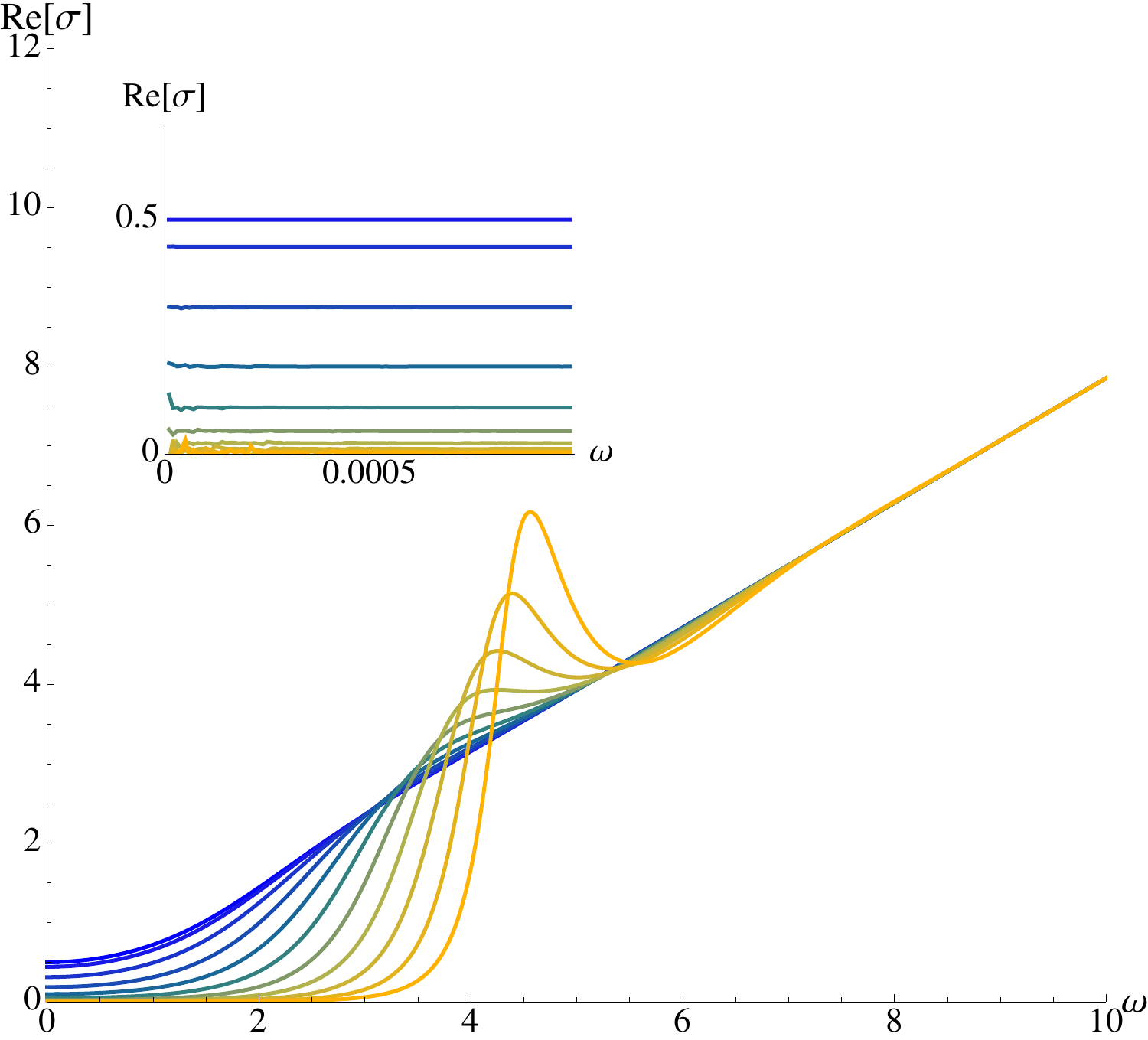}\hspace{2mm}
\includegraphics[width=220pt]{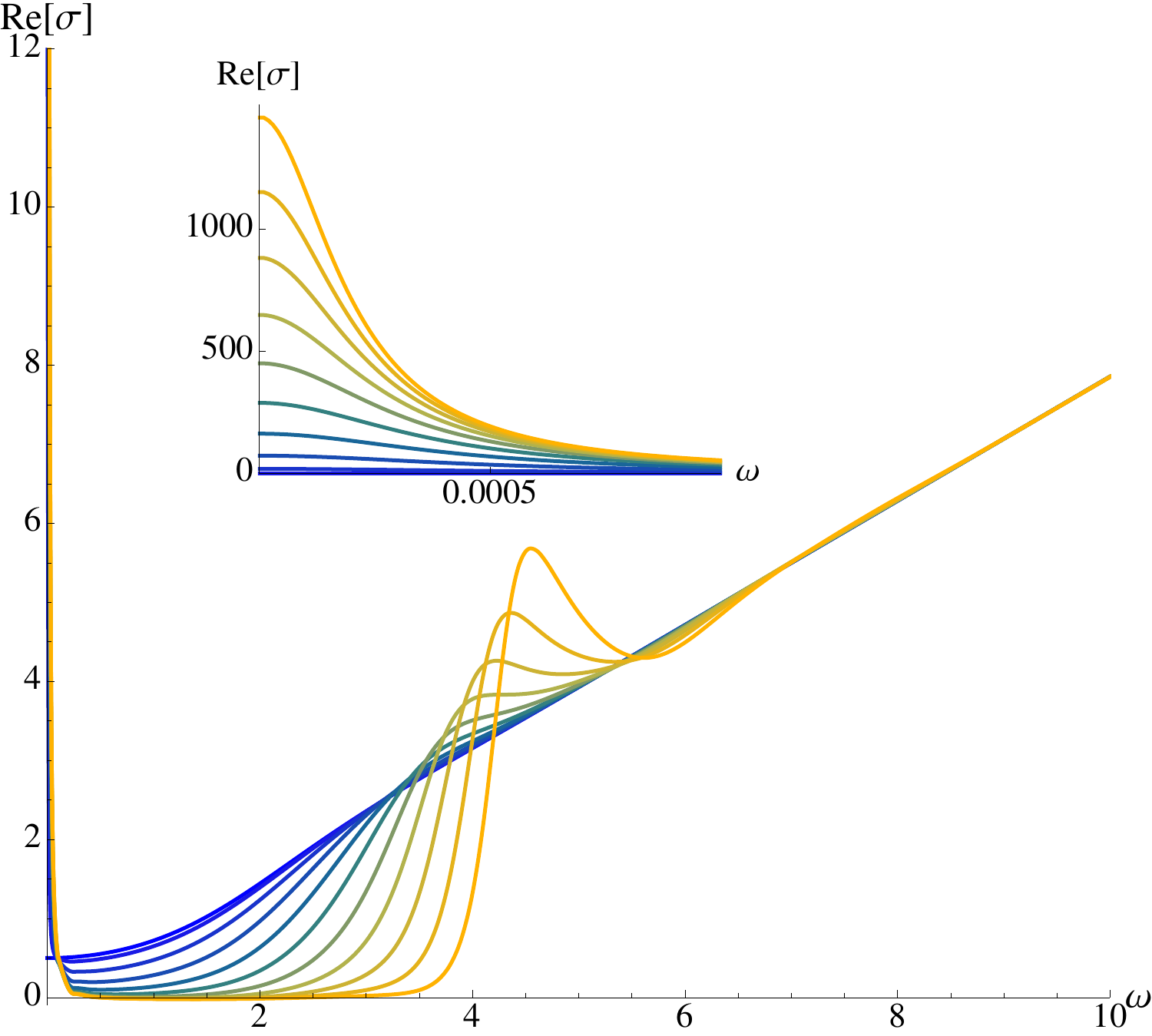}
\caption{\label{fig:relt} Real part of the conductivity in the longitudinal
sector for $\Delta=0$ (Left) and $\Delta=0.1$ (Right). Different colors
correspond to different values of the magnetic field B, from $\kappa B=0$(blue)
to $\kappa B=0.5$ (yellow). The behavior of the conductivity at high
frequencies is qualitatively the same for both values of $\Delta$. The DC
conductivity however shows a Drude peak as soon as the mass ($\Delta$) is
switched on whereas it is a delta function peak centered at $\omega = 0$.}
\end{figure}
\begin{figure}[h] 
\centering
\includegraphics[width=220pt]{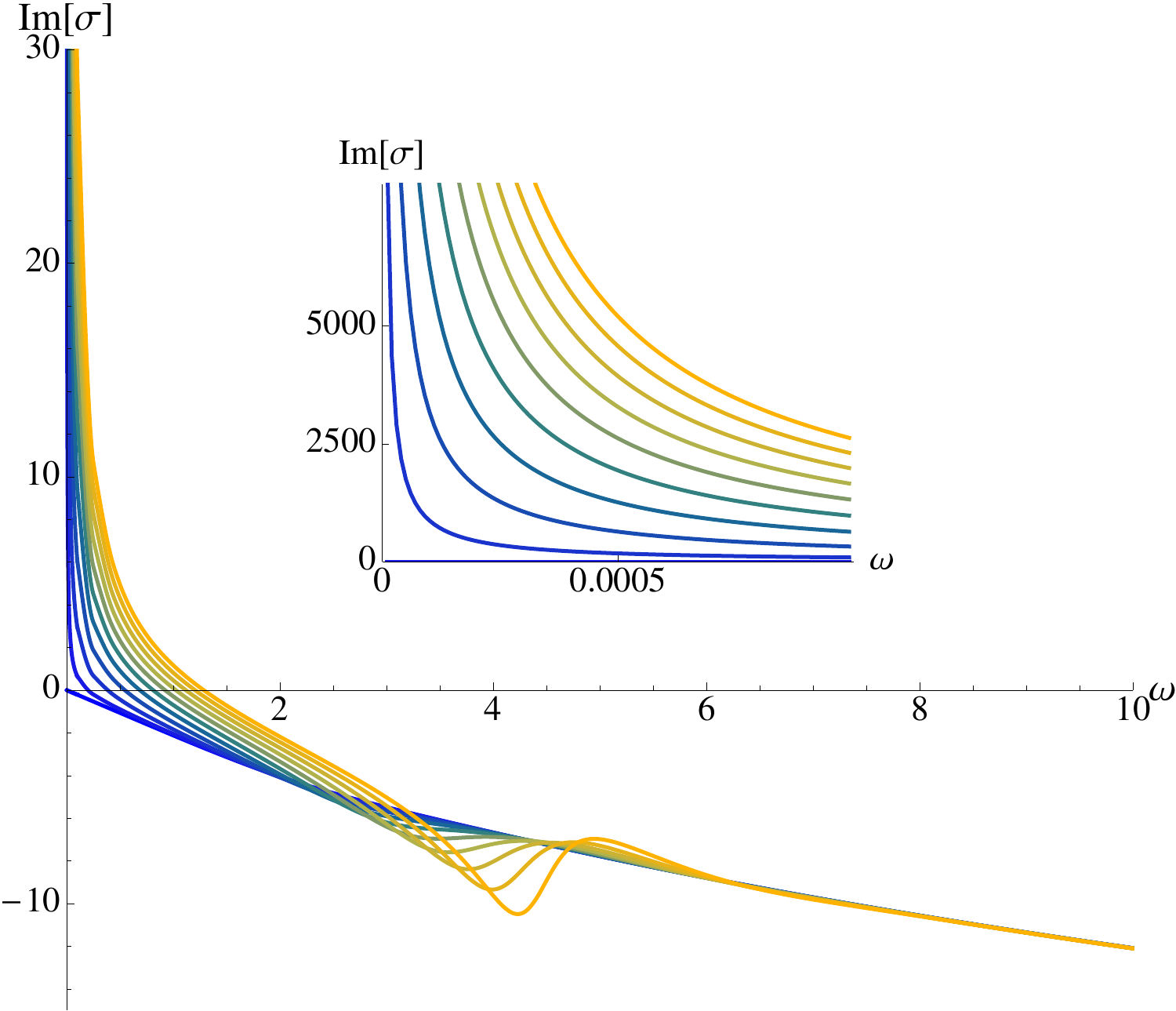}\hspace{2mm}
\includegraphics[width=220pt]{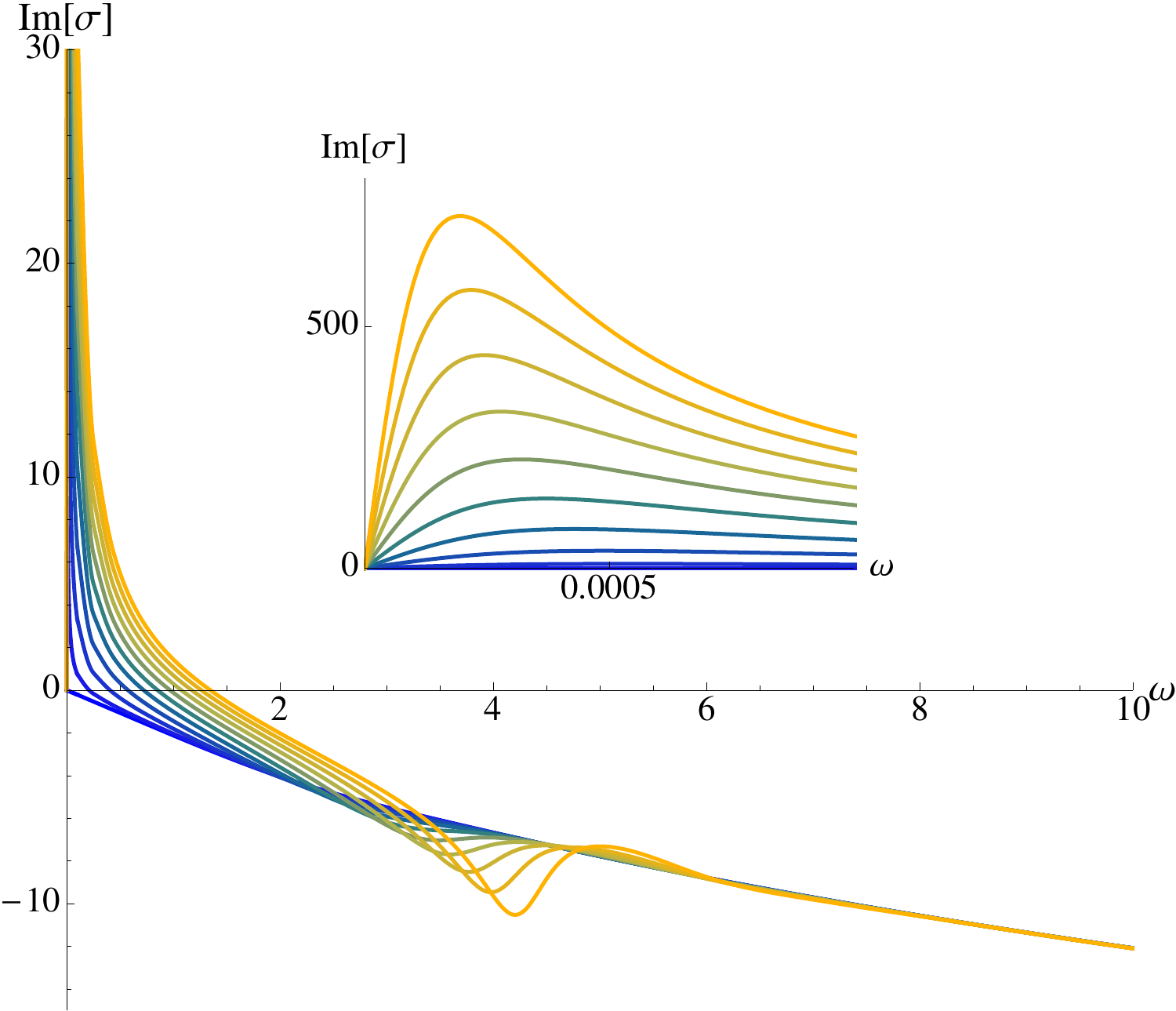}
\caption{\label{fig:imlt} Imaginary part of the conductivity in the longitudinal
sector for $\Delta=0$ (Left) and $\Delta=0.1$ (Right). Different colors
correspond to different values of the magnetic field B, from $\kappa B=0$(blue)
to $\kappa B=0.5$ (yellow). In agreement with the real part in figure
\ref{fig:relt} the zero frequency behavior shows a pole only when the mass is
absent, signaling the presence of a delta function in the real part. As soon as
the mass is switched on $\Im[\sigma]$ vanishes at the origin.}
\end{figure}
\begin{figure}[h] 
\centering
\includegraphics[width=220pt]{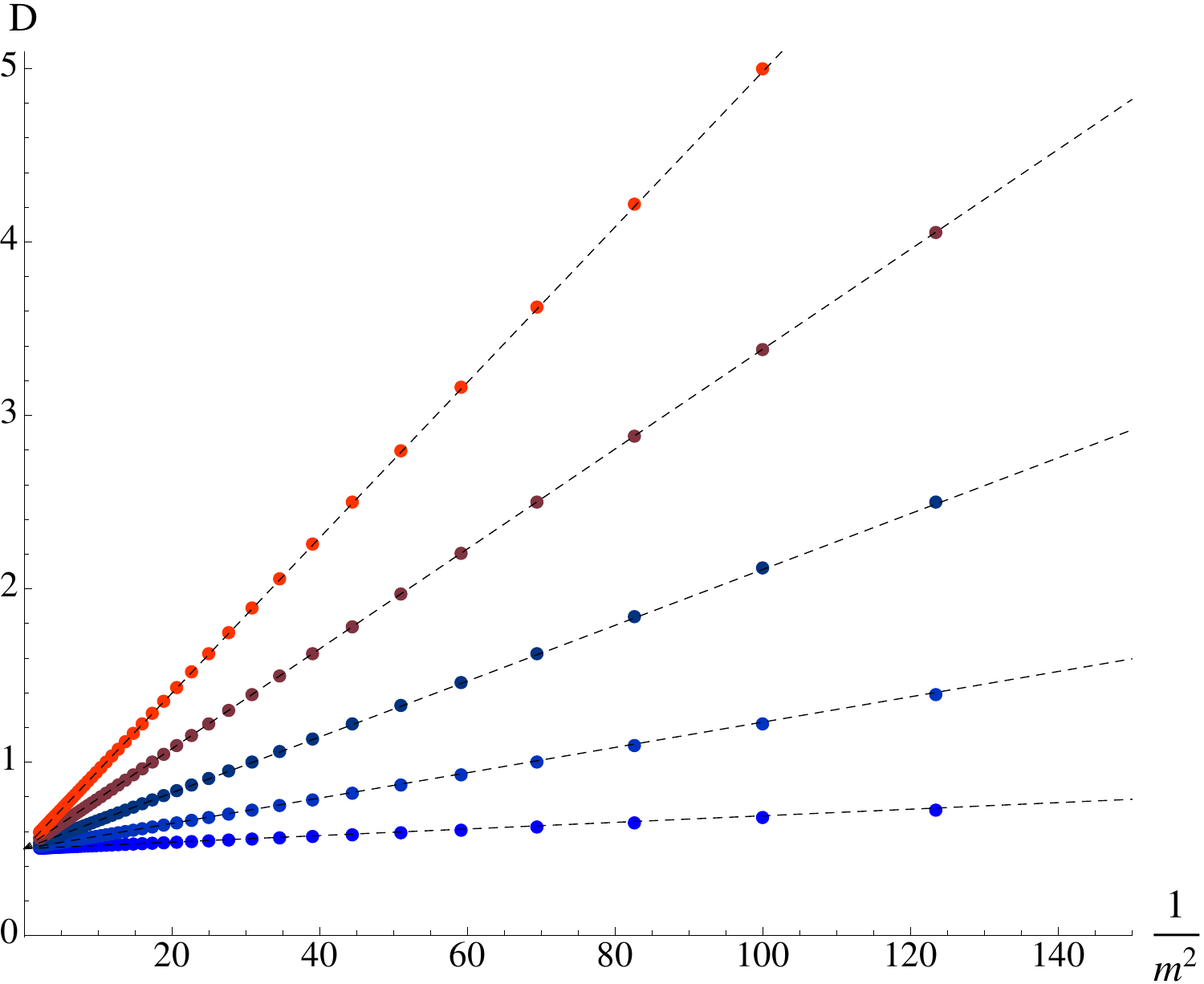}\hspace{2mm}
\includegraphics[width=220pt]{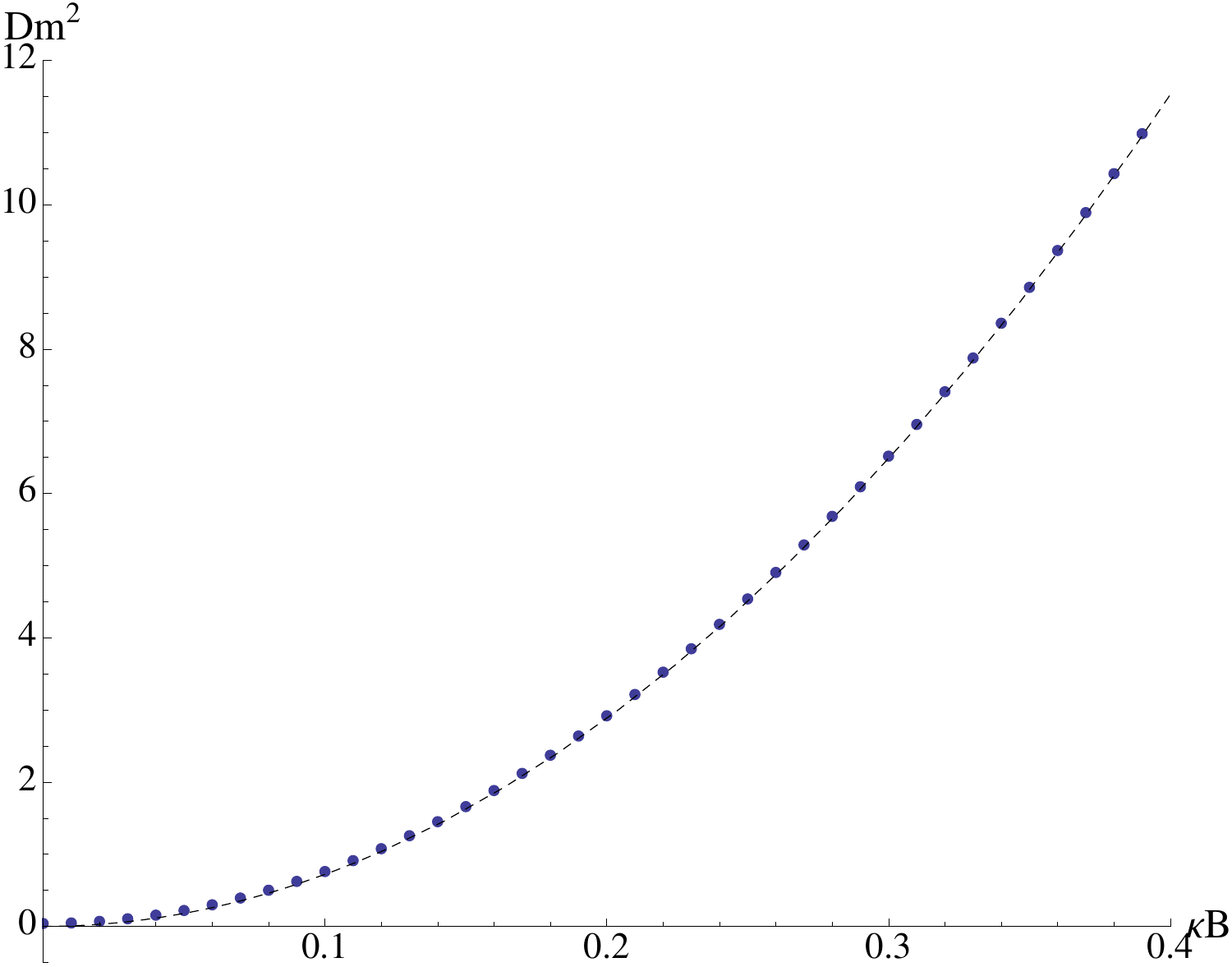}
\caption{\label{fig:condfit} Left: We show the value of the highest point of the
Drude peak (D) against $\frac{1}{m^2}$  for several values of $\kappa B$ from
$\kappa B=0.005$ (Blue) to $\kappa B=0.25$ (Red) . Dashed lines corresponds to
linear fits. Right: Dependence of the slopes in the l.h.s. plot as a function of
$\kappa B$. Dashed line corresponds to a quadratic fit; we find the coefficient
to be $\approx 72$.}
\end{figure}
As a last step we study how the electric conductivity is affected by the mass.
In the absence of mass the CMW induces perfect (i.e. infinite)
conductivities for both the electric and the axial conductivities along the
magnetic field. However, from the QNM analysis of the previous section we know
that this cannot hold anymore. We expect a finite conductivity but with a strong
Drude like peak at zero frequency.
As we will see this is indeed what is happening.\\

In order to analyze the longitudinal conductivity along the magnetic field we
switch on perturbations on top of the background that we used to study the CMW,
namely, an external magnetic field pointing in the $z$-direction. The electric
conductivity along the magnetic field can be extracted from the correlator
\begin{align}
\sigma_{||} = \lim_{\omega\rightarrow 0}\frac{i}{2 \omega} \left<J^z J^z\right>
(\omega, k=0)\,,
%\sigma_{trasnverse} = \lim_{\omega\rightarrow 0}\frac{i}{2 \omega} \left<J^x
%J^x\right> (\omega, k=0)\,.
\end{align}
Since this conductivity is obtained at zero momentum we can assume spatial
homogeneity for the perturbations. The coupled equations of motion can be found
in appendix \ref{app:eomcmw}.
The analysis of the two-point function reveals that for this configuration of the
background the correlator we want to compute has the usual expression
\begin{equation}
\langle J^z J^z \rangle_{\text{Ren.}}= \lim_{r\rightarrow\infty}   r^3
\partial_r {\mathbb{H}} + \omega^2 \log(r)\,.
\end{equation}
We solve the equations numerically with infalling boundary conditions and build
the bulk-to boundary propagator (BBP from now on). Our results 
are shown in figures \ref{fig:relt},\ref{fig:imlt},\ref{fig:condfit}.\\

The well-known Kramers-Kronig relations imply that a pole in the imaginary part
of the conductivity at zero frequency signals
the presence of a delta function peak in the real part, i.e. and infinite DC
conductivity.
As soon as we turn the mass on, we observe that the DC conductivity is not a
delta function anymore (see figures \ref{fig:relt} and \ref{fig:imlt}). This
fact has important consequences in the Ohm's law for an anomalous system with an
explicit breaking term.
It has been first pointed out that the axial anomaly induced a large DC
conductivity in a magnetic field (or a negative magneto resistivity)
in \cite{Nielsen:1983rb}. More recent studies of this phenomenon are
\cite{Gorbar:2013dha,Son:2012bg}.
In these studies Weyl fermions of opposite chirality appear as the effective
electronic excitations at low energies in a 
crystal (Weyl semi-metal). The
associated axial symmetry is however only an approximate one since the
electronic quasiparticles can be scattered from one Weyl cone
into another. The associated scattering rate is called the inter-valley
scattering rate $\tau_i$. It turns out that the conductivity
in these Weyl semi-metals is indeed proportional to the inter-valley scattering
rate. Our findings are in complete analogy, the inverse
of the gap in figure \ref{fig:gama} plays the role of the inter-valley
scattering time leading to a finite but strongly peaked DC magneto conductivity.
\\
% As we have shown in the previous analysis even for a finite mass for the axial
% gauge field in the bulk we get a massless mode in the spectrum due to the
% preserved $U(1)_V$ symmetry. In the presence of the anomaly and a magnetic field
% this mode is (part of) the CMW and induces an infinite DC conductivity parallel
% to the magnetic field direction. Nevertheless, as seen before, it gets altered
% due to the presence of the mass, recovering a diffusive behavior at zero
% momentum (\textcolor{red}{presenting in particular a vanishing residue at
% $\omega=0$}). Indeed, in figure \ref{fig:imlt} one can see how the divergence of
% the Imaginary part at $\omega=0$ vanishes as soon as the mass is switched on,
% signaling the loss of superconductivity. On the other hand, we have seen that
% the mode associated to the diffusive mode of the axial symmetry develops a gap.
% Again, in the region of momenta relevant for the DC conductivity, this mode has
% no linear dependence in the real part of the frequency. The gap has the usual
% impact in the conductivity. As it develops it induces a resonance in the real
% part of the conductivity at zero frequency, as one can see in the left hand side
% of figure \ref{fig:relt}. This resonance disappears as the mass of the bulk
% gauge field increases. One can think of this resonance as the Drude peak coming
% from the delta function in the real part if the bulk mass is turned to zero.  \\
By numerical analysis we find  the dependence of the DC conductivity on $m,
\kappa$ and $B$. Results are shown in In figure \ref{fig:condfit}. We can
approximate it by 
\begin{equation}\label{eq:Bscaling}
D\approx 72\frac{\kappa^2 B^2}{m^2}\,.
\end{equation}
Since in figure \ref{fig:gama} we found that the gap is proportional to $m^2$ we
indeed see that the DC conductivity scale linearly with
the inverse of the Gap as expected. We also find that it depends quadratically
on the magnetic field. Again this is the expected result
at least for small magnetic fields. For larger magnetic fields the weak coupling
analysis shows however a linear dependence on
the magnetic field that can be traced back to the fact that all fermionic
quasiparticles are in the lowest landau level. 

We found that our results show a kind of instability for too large magnetic
field such that we were not able to see this expected cross over
to linear behavior. This might be an artifact of the probe limit or a genuine
instability of the theory at high magnetic fields
(similar to the Chern-Simons term induced instabilities in an electric field
found in \cite{Nakamura:2009tf}). We leave this issue 
for further investigation.

Finally we note that we have checked that the sum rule is fulfilled for several
values of the mass and the magnetic field.
This sum rule takes the form $\frac {d}{dB} \int \Re ( \sigma(\omega)) d\omega =
0$. The sum rule implies that the peak is
built up by shifting spectral weight from higher frequencies into towards
$\omega=0$. In fact this is precisely what can be seen in \ref{fig:relt}
where it is evident that the region of intermediate frequencies gets depleted and
correspondingly a gap in the magneto-optical conductivity
opens up as the magnetic field strength is increased. Note that this gap is
present still in the region where we found quadratic scaling
(\ref{eq:Bscaling}).

 %%%%%%%%%%%%%                   CONCLUSON  %%%%%%%%%%%%%%%%%%%%%%%%%%%%%%%%%

\section{Conclusions}
\label{sec:conclusions}
We have studied anomaly related transport phenomena in a bottom-up holographic
model with massive vector fields and St\"uckelberg axion. 
One of our motivations being that the dynamical part of the axial anomaly, i.e.
the gluonic contribution, is dual to the dynamics of
axions in holography. Its precisely this axion that can be used in the bulk
St\"uckelberg mechanism to give mass to the bulk gauge field. 
The operator dual to this massive gauge field is a non-conserved current and
this non-conservation is manifest in the fact that
we did not find a constraint on the divergence of the current. 
Throughout the paper, we have restricted ourselves to the probe approximation.
\\

Equipped with the above model for a anomalous massive $U(1)$ gauge field, in
section \ref{sec:U1model} we have studied carefully the form of the current
one-point function, showing that the well-known Bardeen-Zumino polynomial does
not exist if the mass $m\ne 0$. The resulting form of the (holographically
renormalized) current tends to the \emph{consistent} definition in the massless
limit. Moreover, as described by (\ref{dcurrent}) the divergence of such a
current is not constrained. Moving to the two-point functions, the anomalous
conductivity $\sigma_{55}$ has been computed using its definition via a Kubo
formula. We find that its value corresponds to the one associated to to the 
consistent current in the zero mass limit. We also showed that the QNM spectrum
has a gap in contrast to the massless case in which
there exists a hydrodynamic diffusion mode. We stress that the non-conserved
current $J^i$ is not a hydrodynamic variable because
of this gap. Furthermore the parameter $\mu_5$ is not a chemical potential but a
coupling constant. Nevertheless we think it would be
an interesting exercise to work out constitutive relations of for the
non-conserved current extending the well-established methods of
the fluid/gravity correspondence \cite{Bhattacharyya:2008jc} to this case. 

%  Let us stress that one has to be very careful when trying to interpret our
% results for $\sigma_{55}$ from the hydrodynamic point of view. In this context,
% the main difference between the global anomaly and the dynamical one is that the
% latter does not depend on external non-dynamical sources, but on dynamical
% (gluon) fields that will be responsible for a non-negligible contribution to
% $\partial_i\left< J^i_5\right>$ through the anomaly, that is, the axial current
% is by all means a non-conserved current. This in particular implies that $J^i_5$
% will not be a hydrodynamic quantity and thus will not enter the hydrodynamic
% constitutive relations. Furthermore, the source $\mu_5$ cannot be regarded as a
% hydrodynamic parameter. In addition, the gauge field(s) do enter the equations
% of (magneto)hydrodynamics, even though in our holographic approach the gauge
% sector cannot be studied. In fact, having a non-conserved axial symmetry most
% likely implies that $(A_i -\partial_i\theta ); \ i=x,y,z$ is a licit
% gauge-invariant source whose effects can be analyzed. This suggest that an
% immediate extension of our work is to allow for the above quantity to be
% different from zero and study the consequences of it. Afterwards, in order to
% clarify the non-conservation of the current, we studied the QNM spectrum,
% focusing in the (pseudo-)diffusive mode. As expected, it develops a gap when the
% mass is increased from zero, indicating that the current is no longer conserved
% if $m>0$.\\

In section \ref{sec:U1U1model} we implemented the interplay between
non-conserved axial and conserved vector currents.
We also studied a  wider set of anomalous conductivities using Kubo formulae
(\ref{kubo}), (\ref{kubos}) and (\ref{kuboss})
We found that (as expected) the axial conductivity is identical to the case
with only one axial gauge field.
The Chiral Separation conductivity is independent of the source $\mu_5$, behaves
linearly with $\mu$ and increases with the mass, as depicted in figure
\ref{fig:cse}. Finally, the Chiral Magnetic conductivity vanishes for all the
values of $m$ that we have studied; we interpret this fact as an effect of the
source that ensures that the background is time-independent. As $m\rightarrow
0$, all the conductivities approach the value corresponding to consistent
definition of the currents. Subsection \ref{subsec:CMW} is devoted to the study
of the Chiral Magnetic Wave (CMW) \cite{Kharzeev:2010gd} in the presence of
mass. First, we perform an analytic analysis of the type of modes in a
phenomenological model
that implements the axial non-conservation via a relaxation term (see equation
(\ref{eq:noncons})). 
This model predicts that a propagating wave like mode can build up only for
large enough momentum. Indeed we find from our quasinormal
mode analysis that the model can be fitted very well to the QNM spectrum and
that indeed a propagating chiral magnetic wave is
absent for small momenta.

Finally we have also studied the negative magneto resistivity and showed that a
sum rule holds for the magneto-optical conductivity.
The strength of the DC conductivity is proportional to the square of the
magnetic field and inverse proportional to the gap.
This is in agreement with weak coupling considerations for small magnetic fields
and an inter-valley scattering relaxation time 
for axial charge. Unfortunately we were not able to see the expected cross-over
to linear behavior in the magnetic field because
our numerics indicated a possible instability at large B-field. If this is an
artifact of the probe limit (which assumes negligible 
backreation of the gauge field on the geometry) or a genuine instability we
leave to further investigation.

% % We depict in figure \ref{fig:REIM} the outcome of our subsequent numerical
% computation in the holographic model, showing that if $m\ne 0$ there exists a
% critical momentum $k_C$, given by (\ref{eq:critk}), below which there exist one
% massless (related to the $U(1)_V$ symmetry) and one massive mode in the spectrum
% of QNMs that are purely imaginary (i.e. $\Re[\omega]=0$). For $k>k_C$ we obtain
% $\Re[\omega] \ne 0$, recovering a CMW-like mode that is however gapped. At
% $m=0$, we recover the usual dispersion relation for the CMW; in particular, it
% becomes ungapped. These numerical results are almost identical to
% (\ref{eq:wpmconc}), see figure \ref{fig:REIM}. Finally, we also study how the
% above behaviour of the CMW when the mass is swotched on affects the electric
% conductivity. It is well-known that the electric DC conductivity is inifnite if
% $m=0$. After turning the mass on, $\Re[\sigma(k=0)]$ develops a drude peak,
% whereas $\Im[\sigma(k=0)] =0$. The plots correspond to figures \ref{fig:relt}
% and \ref{fig:imlt}. This is a consequence of the fact that at $k=0$ the lowest
% QNMs feature no real part (green line of figure \ref{fig:REIM}) : one of them is
% massless \textcolor{red}{but its residue vanishes at zero momentum}, whereas the
% other one is gapped. We carry out an analysis of the height of the Drude peak (D
% in the text) summarized in figure \ref{fig:condfit}; we can approximate the
% numerical outcome by
% \begin{align}
% D\approx 72\frac{\kappa^2 B^2}{m^2}\,.
% \end{align}

This brings us to possible generalizations of the present work. First we would
like to mention that the usage of St\"uckelberg axions in
the context of holographic studies of anomaly induced transport has recently
also been suggested in \cite{Gursoy:2014ela}.
Our model is a first step into this direction and following
\cite{Gursoy:2014ela} one might improve it by giving up conformal symmetry
by working directly with the model of \cite{Klebanov:2002gr} or a suitable
simplification thereof.
Another rather straightforward generalization would be to take the backreaction
onto the geometry into account. This opens the way to
study also the generalizations of the chiral vortical effect and one could also
include the mixed gauge gravitation anomaly. 
As we emphasized the non-conserved currents do not strictly belong to the set
of hydrodynamic variables. But one can easily imagine
a situation in which the gap of the lowest quasinormal mode in the massive
vector sector is much smaller than the separation to 
the higher QNMs. In this case it would make sense to include these modes in the
gauge/gravity correspondence and work out the constitutive
relations. A very interesting question arises in connection to the to possibility
of defining covariant or consistent currents in the massless case. We found that
the non-conserved current goes over into the consistent current in the zero mass
limit. Is is possible to generalize the notion of covariant current to the massive
case? It is also known that the consistent currents are not unique but can be 
redefined by adding finite counterterms (the Bardeen counterterms). Its precisely this
choice of counterterms that allows to shift the anomaly completely into the 
axial sector (even when a mixed gravitational anomaly is present). It would be
certainly interesting to include the gravitational anomaly and to see how the
Bardeen-Zumino terms arise in the zero mass limit.

% As future directions, the most straightforward extension would be that of
% considering sources for the spatial components of the gauge field, as mentioned
% earlier. Moreover, one can construct more involved models presenting
% St\"uckelberg-like terms (see  for a discussion on this issue). In fact, it is
% known that the gluonic sector plays a major role for the anomalous response of
% the Quark Gloun Plasma \cite{Kharzeev:2004ey,Kharzeev:2007tn,Kharzeev:2007jp}.
% Therefore, we believe that the models studied here represent the correct
% framework in which to analyze the implications of chiral transport to Heavy Ion
% collisions. In this sense, it would be phenomenologically interesting to examine
% the consequences of embedding our toy model into a more realistic (i.e QCD-like)
% set-up.
                                                                                
%%%%%%%%%%%%%%%%%%        ACKNOWLEDGEMENTS                %%%%%%%%%%%%%%%%%%%%

\section*{Acknowledgments}
We would like to thank Ioannis Papadimitriou, Umut Gursoy and Claudio Coriano for useful discussions and
feedback. \\
This work has been supported by Plan Nacional de
Altas Energ\'\i as FPA2009-07890, Consolider Ingenio 2010 CPAN CSD200-00042 and
Severo Ochoa award SEV-2012-0249. L.M. has been supported by FPI-fellowship
BES-2010-041571. A. J. has been supported by FPU fellowship AP2010-5686. 

 \newpage                                                                                  
%%%%%%%%%%%%%%%%%%%                      APPENDIX         %%%%%%%%%%%%%%%%%%%%%%

\begin{appendix}
\section{Holographic Renormalization}\label{app:ren}
\subsection{U(1) Model}

In order to renormalize the theory shown in (\ref{eq:1}) we follow the procedure
in \cite{Papadimitriou:2004ap}. Within this approach the renormalization
procedure consists of an expansion of the canonical momenta and the On-Shell
action $\lambda$ in eigenfunctions of the dilatation operator. This operator can
be obtained taking the asymptotic leading term of the radial derivative

\begin{align}
\partial_r= \int d^dx \left(  \dot \gamma_{ij} \frac{\delta}{\delta
\gamma_{ij}}+   \dot A_{i} \frac{\delta}{\delta A_{i}}+  \dot \theta
\frac{\delta}{\delta \theta}  \right) \sim  \int d^dx \left(  2 \gamma_{ij}
\frac{\delta}{\delta \gamma_{ij}}+   \Delta A_{i} \frac{\delta}{\delta A_{i}}+ 
O(e^{-r}) \right) 
\end{align}
\begin{equation}\label{dilatator}
\delta_D=\int d^dx \left(  2 \gamma_{ij} \frac{\delta}{\delta \gamma_{ij}}+
\Delta A_{i} \frac{\delta}{\delta A_{i}}\right) 
\end{equation}
Notice that this operator is not gauge invariant. Nevertheless $S_{C.T.}$ must
be gauge invariant since the bulk Lagrangian is invariant too. Therefore
$S_{C.T.}$ must be expressible as a functional of $B_i\equiv
A_i-\partial_i\theta$. We will see that this is indeed the case even though we
expand the on-shell action in eigenfunctions of the (non gauge invariant)
dilatation operator $\delta_D$.  We choose the axial gauge $A_r=0$. Recall that
\begin{equation}
\Delta(\Delta+2)=m^2.
\end{equation}
Our notation for the eigenfunctions of the dilatation operator reads
\begin{align}
\delta_D X_{(a)}= -a X_{(a)}\hspace{2cm}  \delta_D X_{(4)}=-4 X_{(4)}-2\tilde
X_{(4)}
\end{align}
All our results were obtained in the probe limit and therefore, for simplicity,
we adapt the renormalization procedure to this limit. This implies that we will
use the e.o.m. for the fields, instead of the Hamiltonian constraint in Einstein
equations, to determine the eigenfunctions of the dilatation operator the
canonical momenta are expanded in. In addition we set the extrinsic curvature
$K_{ij}\equiv \dot \gamma_{ij}=2\gamma_{ij}$, which in our setup is enough for
the boundary analysis.
The matter e.o.m., written in terms of $E_i\equiv \dot A_i$ and $\Pi \equiv \dot
\theta$ are:

\begin{align}
&\dot E_i + 2 E_i -m^2\left( A_i -\partial_i \theta\right)+\partial^j
F_{ji}+2\kappa \epsilon^{irjkl} E_j F_{kl}=0\,,\label{eqe}\\
&\dot \Pi + 4\Pi - \partial^i \left( A_i - \partial_i \theta \right)=0\,,\\
&\Pi= \frac{1}{m^2}\left(\partial^i E_i -\kappa \epsilon^{rijkl}
F_{ij}F_{kl}\right)\,.\label{eqpi}
\end{align}
 
With (\ref{dilatator}) and the e.o.m. we can determine the explicit form
of the different terms in the expansions
\begin{align}
&E_i=E_{i(-\Delta)} +E_{i(0)} +E_{i(2-2\Delta)} +E_{i(2-\Delta)}
+E_{i(2)}+...\,,\\
&\Pi=\Pi_{i(2-\Delta)} +\Pi_{i(2)} +...
\end{align}
\begin{align}
&E_{i(-\Delta)}=\Delta A_i\,, \nonumber\\
&E_{i(0)}=-\Delta \partial_i \theta\,,\nonumber\\
&\Pi_{(2-\Delta)}=\frac{1}{(\Delta+2)}  \partial_i A^i\,,\nonumber\\ 
&\Pi_{(2)}=\frac{-1}{(\Delta+2)} \Box \theta\,.\\
\end{align}

Other terms like $E_{i(2-2\Delta)}$ are non-zero but as we will see they not contribute to the counterterms.
We can determine the expressions for the higher order operators needed to
expand the radial derivative:
\begin{align}
\partial_r =
\delta_D+\delta_{(\Delta)}+\delta_{(2-2\Delta)}+\delta_{(2-\Delta)}+\delta_{(2)}
+\delta_{(2+\Delta)}+...
\end{align}
\begin{align}
&\delta_{(\Delta)}= \int d^dx' E_{i(0)}(x')\frac{\delta}{\delta
A_i(x')}\nonumber\\
&\delta_{(2-\Delta)}= \int d^dx'\left(E_{i(2-2\Delta)}(x')\frac{\delta}{\delta
A_i(x')}+\Pi_{(2-\Delta)}(x')\frac{\delta}{\delta
\theta(x')}\right)\nonumber\\
&\delta_{(2)}= \int d^dx'\left( E_{i(2-\Delta)}(x')\frac{\delta}{\delta
A_i(x')}+ \Pi_{(2)}(x')\frac{\delta}{\delta \theta(x')}\right)\nonumber\\
&\delta_{(2+\Delta)}= \int d^dx' E_{i(2)}(x')\frac{\delta}{\delta A_i(x')}
\end{align}

Once we have these we just need the equation for the On-Shell action
\begin{align}
 \dot \lambda + \lambda - \mathcal{L}_m=0
\end{align}
\begin{align}\label{eqosa}
\dot \lambda +4\lambda +\frac{1}{2}E_iE^i+\frac{m^2}{2}\Pi^2 +
\frac{m^2}{2}(A_iA^i-2A_i \partial^i\theta +\partial_i\theta \partial^i
\theta)+\nonumber\\
\frac{1}{4}F_{ij}F^{ij}+\frac{4\kappa}{3}(A_i-\partial_i\theta)E_j
F_{kl}\epsilon^{irjkl} -\frac{\kappa}{3}\Pi F_{ij}F_{kl}\epsilon^{irjkl} =0
\end{align}

To determine the terms of the eigenfunction expansion of the On-Shell action
\begin{align}
\lambda=&\lambda_{(0)}+\lambda_{(2-2\Delta)}+\lambda_{(2-\Delta)} 
+\lambda_{(2)} +\lambda_{(4-4\Delta)} +\nonumber\\
& \lambda_{(4-3\Delta)}+\lambda_{(4-2\Delta)} +\lambda_{(4-\Delta)}
+\lambda_{(4)} + \tilde\lambda_{(4)} \log e^{2r} +... 
\end{align}

It is important to remark that depending on the value of $0\leq\Delta\leq1$ new
terms may appear in this expansion. For example, the next possible term in this
expansion is $\lambda_{(6-6\Delta)}$. Therefore, as in the rest of the paper, we
restrict our analysis to 
\begin{equation}\label{condicion}
6-6\Delta>4 \rightarrow\Delta <\frac{1}{3}.
\end{equation}
Furthermore, for a large enough mass $(\Delta=1)$ the number of possible
counterterms becomes infinite. This is to be expected since for such a value of
the mass the operator dual to the gauge field becomes marginal.\\

%The following identities will be usefull :
%\begin{align}
%&\delta_D A_i=\Delta A_i \hspace{2cm} \delta_D \theta=0 \hspace{2cm} \delta_D
%\gamma_{ij}=2 \gamma_{ij}\\&\delta_D \gamma^{ij}=-2 \gamma^{ij}   \hspace{1cm} 
%\delta_D \epsilon^{\alpha\beta\mu\nu\rho}\equiv\delta_D
%\frac{\epsilon(\alpha\beta\mu\nu\rho)}{\sqrt[]{-g}} =-4
%\epsilon^{\alpha\beta\mu\nu\rho}
%\end{align}

We are now ready to proceed solving (\ref{eqosa}) order by order in dilatation
weight
\begin{align}
 &\lambda_{(0)}\,\,\,\,\,\,\,\,\,\,\,=0\\
 &\lambda_{(2-2\Delta)}\,\,=\frac{-\Delta}{2}A_{i}A^{i}\\
&\lambda_{(2-\Delta)}\,\,\,\,=\Delta\partial_i \theta A^i\\
&\lambda_{(2)}\,\,\,\,\,\,\,\,\,\,\,=\frac{-\Delta}{2}
\partial_i\theta\partial^i\theta
\end{align}

At this point one can see that these first terms of the O.S. action expansion
can be rearranged in terms of the $B_i$ field:

\begin{equation}\label{eqginv}
\lambda_{(2-2\Delta)}+\lambda_{(2-\Delta)}+\lambda_{(2)}=
-\frac{\Delta}{2}B_{i}B^{i}
\end{equation}

It is a nice check to find all the terms explicitly and then rearrange them like
this although, as mentioned before, this is to be expected. Moreover we can use
this in our advantage: once one obtains a counterterm which is only proportional
to $A_i$ the following terms can be determined by just imposing that $\lambda$
has to be gauge invariant.
\begin{align}
&\lambda_{(4-4\Delta)}=0
&\lambda_{(4-3\Delta)}=0
\end{align}

Let us analyze the following term with some detail
\begin{align}\label{necesario}
&\left. \dot \lambda
\right|_{(4-2\Delta)}+4\lambda_{(4-2\Delta)}+E_{i(-\Delta)}E^i_{(2-\Delta)}
+\frac{m^2}{2}\Pi_{(2-\Delta)}^2+\frac{1}{4}F_{ij}F^{ij}+
\frac{4\kappa}{3}\epsilon^{rijkl}
F_{jk}\left(E_{i(0)}A_l-E_{i(-\Delta)}\partial_l\theta\right)=0\nonumber\\
&(\delta_D+4) \lambda_{(4-2\Delta)} +\delta_{(2-\Delta)}
\lambda_{(2-\Delta)}+\delta_{(2)}
\lambda_{(2-2\Delta)}+E_{i(-\Delta)}E^i_{(2-\Delta)}+\frac{m^2}{2}\Pi_{
(2-\Delta)}^2+\frac{1}{4}F_{ij}F^{ij}=0\nonumber\\
&\lambda_{(4-2\Delta)}=\frac{1}{4(\Delta+2)}\partial_i A^i \partial^j A_j
-\frac{1}{8\Delta} F_{ij}F^{ij}    
\end{align}

It is remarkable that the term proportional to $\kappa$ vanishes due to the
contraction of a symmetric
$\left(E_{i(0)}A_l-E_{i(-\Delta)}\partial_l\theta\right)$ and an antisymmetric
$\epsilon^{rijkl}$ tensor. Here we see the importance of the coupling of the
St\"uckelberg filed to the Chern-Simons term. If we had not added it, at this
point we would have found an extra counter term of the form $-\theta\wedge F
\wedge F$.  In this expression we have neglected total derivatives. From this
last equation we can infer the following two orders by imposing gauge
invariance. So, in terms of the gauge invariant field $B_i$ the counterterm
reads:

\begin{align}
\lambda_{4-2\Delta}+\lambda_{4-\Delta}+\lambda_{4}^*=\frac{1}{4(\Delta+2)}
\partial_i B^i \partial_j B^j -\frac{1}{8\Delta} F_{ij}F^{ij}   
\end{align}
Note that we cannot determine $\lambda_{(4)}$ with just the boundary analysis.
$\lambda_{4}^*$ is just a part of $\lambda_{4}$ which is imposed by gauge
invariance and that can be obtained from the asymptotics.

We only lack the $\sim \log$ term, that is obtained by evaluating the equation
to 4th order
\begin{align}
\tilde\lambda_{(4)}=0
\end{align}
Thus, the $S_{CT}$ reads:
\begin{equation}
S_{CT}=\int_\partial d^dx\,\, \sqrt[]{-\gamma}\left( \frac{\Delta}{2}B_{i}B^{i}
- \frac{1}{4(\Delta+2)}\partial_i B^i \partial_j B^j  +
\frac{1}{8\Delta}F_{ij}F^{ij}       \right)
\end{equation} 

\subsection{U(1)xU(1)  model} \label{app:ren2}
 Few things change if we introduce a second gauge field (non-massive,
non-anomalous in the boundary). The asymptotic behavior remains unchanged.
Specially, the vector gauge field behaves as it usually does and thus it does
not contribute to the dilatation operator.
\begin{equation}
  \delta_D=\int d^dx \left(  2 \gamma_{ij} \frac{\delta}{\delta \gamma_{ij}}+  
\Delta A_{i} \frac{\delta}{\delta A_{i}}+  O(e^{-r}) \right)
\end{equation}

 The equation of the O.S. action has to be modified:
\begin{align}
&\dot \lambda +4\lambda +\frac{1}{2}E_iE^i
+\frac{1}{2}\Sigma_i\Sigma^i+\frac{m^2}{2}\Pi^2 + \frac{m^2}{2}(A_iA^i-2A_i
\partial^i\theta +\partial_i\theta \partial^i \theta)+\nonumber\\
&\frac{1}{4}F_{ij}F^{ij}+\frac{1}{4}H_{ij}H^{ij}+2\kappa(A_i-\partial_i\theta)
\left( E_j F_{kl}+3\Sigma_j H_{kl} \right) \epsilon^{irjkl} -\nonumber\\
&\frac{\kappa}{2}\Pi \left(F_{ij}F_{kl} +3H_{ij}H_{kl}\right)\epsilon^{irjkl} =0
\end{align}

Where $\Sigma$ and $H_{ij}$ are the the momentum\footnote{as we did with the
axial field we define $\Sigma_i\equiv \dot V_i$.} and the field strength of the
vector gauge field.
It is not difficult to realize that the only term proportional to $V_i$ that
will contribute to the divergent part of $\lambda$ is the kinetic term
$H_{ij}H^{ij}$. Since this is of order 4, it will only contribute to the
logarithmic term and therefore:

\begin{equation}
S_{CT}=\int_\partial d^dx\,\, \sqrt[]{-\gamma}\left( \frac{\Delta}{2}B_{i}B^{i}
- \frac{1}{4(\Delta+2)}\partial_i B^i \partial_j B^j  +
\frac{1}{8\Delta}F_{ij}F^{ij} +\frac{1}{8}H_{ij}H^{ij}  \log e^{2r}     \right)
\end{equation}

%%%%%%%%%%%%%%%%%%%%%%%%%%%%%%%%%%%%%%%%%%%%%%%%%%%%%%%%%%%%%%%%%%%%%%%%%%%%%%%%
%%%%%%%%%%%%%%%%%%%%%%%%%%%%%%%%%%%%%%%%%%%%%%%%%%%%%%%%%%%%%%%%%%%%%%%%%%%%%%%%
%%%%%%%%%%%%%%%%%%%%%%%%%%%%%%%%%%%%%%%%
%%%%%%%%%%%%%%%%%%%%%%%%%%%%%%%%%%%%%%%%%%%%%%%%%%%%%%%%%%%%%%%%%%%%%%%%%%%%%%%%
%%%%%%%%%%%%%%%%%%%%%%%%%%%%%%%%%%%%%%%%%%%%%%%%%%%%%%%%%%%%%%%%%%%%%%%%%%%%%%%%
%%%%%%%%%%%%%%%%%%%%%%%%%%%%%%%%%%%%%%%%
%%%%%%%%%%%%%%%%%%%%%%%%%%%%%%%%%%%%%%%%%%%%%%%%%%%%%%%%%%%%%%%%%%%%%%%%%%%%%%%%
%%%%%%%%%%%%%%%%%%%%%%%%%%%%%%%%%%%%%%%%%%%%%%%%%%%%%%%%%%%%%%%%%%%%%%%%%%%%%%%%
%%%%%%%%%%%%%%%%%%%%%%%%%%%%%%%%%%%%%%%%
                                                                                
                          %%%%%%%%%%%%%%%%%%%%%%%%%%% 
%%%%%%%%%%%%%%%%%%%%%%%%%%%%%%%%%%%%%%%%%%%%%%%%%%            U1 n-point
functions      %%%%%%%%%%%%%%%%%%%%%%%%%%%%%%%%%%%%%%%%
                                                                                
                          %%%%%%%%%%%%%%%%%%%%%%%%%%%
%%%%%%%%%%%%%%%%%%%%%%%%%%%%%%%%%%%%%%%%%%%%%%%%%%%%%%%%%%%%%%%%%%%%%%%%%%%%%%%%
%%%%%%%%%%%%%%%%%%%%%%%%%%%%%%%%%%%%%%%%%%%%%%%%%%%%%%%%%%%%%%%%%%%%%%%%%%%%%%%%
%%%%%%%%%%%%%%%%%%%%%%%%%%%%%%%%%%%%%%%%
%%%%%%%%%%%%%%%%%%%%%%%%%%%%%%%%%%%%%%%%%%%%%%%%%%%%%%%%%%%%%%%%%%%%%%%%%%%%%%%%
%%%%%%%%%%%%%%%%%%%%%%%%%%%%%%%%%%%%%%%%%%%%%%%%%%%%%%%%%%%%%%%%%%%%%%%%%%%%%%%%
%%%%%%%%%%%%%%%%%%%%%%%%%%%%%%%%%%%%%%%%
%%%%%%%%%%%%%%%%%%%%%%%%%%%%%%%%%%%%%%%%%%%%%%%%%%%%%%%%%%%%%%%%%%%%%%%%%%%%%%%%
%%%%%%%%%%%%%%%%%%%%%%%%%%%%%%%%%%%%%%%%%%%%%%%%%%%%%%%%%%%%%%%%%%%%%%%%%%%%%%%%
%%%%%%%%%%%%%%%%%%%%%%%%%%%%%%%%%%%%%%%%

\section{Correlators in the U(1) model}

\subsection{1-point function}\label{app:1pfu1}

In order to derive the 1-point function of the (non-conserved) vector operator
dual to the gauge field we write fields as background plus perturbations
\begin{equation}
\mathcal{A}_\mu = A_\mu + a_\mu \hspace{3cm}  \theta= \theta + \phi
\end{equation}
we expand the \textbf{renormalized} action to first order in the perturbations
\begin{align}\label{1order}
 S^{(1)}_{\mathcal{R}}= &\int dr\,d^4x\,\, \sqrt[]{-g} \left[  a_\mu
\left(\nabla_\nu F^{\nu \mu} -m^2 (A^\mu-\partial^\mu \theta)+ \kappa
\epsilon^{\mu \alpha \beta \gamma \rho} F_{\alpha \beta} F_{\gamma \rho}  
\right)-\phi  \nabla_\mu \left( A^\mu -\partial^\mu \theta   \right)  \right]
+\nonumber\\
& \int_\partial d^4x  \,\, \sqrt[]{-g} \left[ \,a_i \left ( F^{i r}+\frac{4}{3}
\kappa (A_j-\partial_j \theta) F_{kl} \epsilon^{r i j k l} \right)  - \phi
(F_{ij} F_{kl} \epsilon^{r i j k l}+m^2(A^r-\partial^r\theta))
\right]+\nonumber\\
& \int_\partial d^4x  \,\, \sqrt[]{-\gamma} \,\,a_i \left (    \Delta
(A^i-\partial^i\theta) + \frac{1}{2(\Delta+2)}   \partial^i (\partial_j A^j -
\Box \theta) -\frac{1}{2\Delta} \partial_j F^{j i}   \right)  + \nonumber\\
& \int_\partial d^4x  \,\, \sqrt[]{-\gamma} \,\,\phi    \left(   \Delta 
(\partial_j A^j - \Box \theta)   + \frac{1}{2(\Delta+2)} \Box (\partial_j A^j -
\Box \theta)    \right)   
\end{align}
The bulk integral contains the e.o.m. for the background fields. The second line
shows the boundary term that arises from the unrenormalized action $S$ whereas
the third and fourth lines contain the expansion of the counter term
action$S_{CT}$. By inspection of the equations of motion one finds that the most
general asymptotic expansion of the fields reads
\begin{align}\label{expansion}
A_\mu\sim  &\sum_{i=0}^\infty A_{\mu(i)} r^{\Delta-i} + \sum_{i=0}^\infty \tilde
A_{\mu(i)} r^{-2-\Delta-i }+ \sum_{i=0}^\infty \tilde \theta_{\mu(i)} r^{-i }
+\nonumber\\
& \sum_{n>1,i\geq 2(n-1)}^\infty \omega_{\mu(n,i)} r^{n\Delta-i}+
\sum_{n>1,i\geq 3n}^\infty  \tilde\omega_{\mu(n,i)} r^{-n\Delta-i}
+\sum_{i\geq4} A_{L(i)}r^{(-i)}\log(r)\\\nonumber\\
\theta \sim &\sum_i \theta_{(i)}r^{(-i)}+  \sum_{n\geq1, i\geq 2n} \tilde
\Psi_{(n,i)}r^{(n\Delta-i)} +  \sum_{n\geq1, i\geq 3n+2} \tilde
\Psi_{(-n,i)}r^{(-n\Delta-i)} +\nonumber\\
&\sum_{i\geq4} \theta_{L(i)}r^{(-i)}\log(r).
\end{align}
With $\Delta = \sqrt[]{1+m}-1$ that is bounded to be $\Delta<1$. The coefficient
of the leading (non-normalizable mode) term $A_{(0)x}$ is to be identified with
the source of the dual operator. $\tilde A_{(0)x}$ is the coefficient of the
normalizable mode. $\omega_{(n,i)}, \tilde \omega _{(n,i)}$ arise due to the
non-linearities of the e.o.m. and can be expressed as functionals of the sources
of the ``other'' components of the gauge field $A_{(0)y\neq x}$. Finally, the
$\tilde \theta$ and $A_L$ terms arise from the coupling to the St\"uckelberg
field and are functionals of the source of $\theta$; the logarithmic terms are
sub leading w.r.t. the normalizable mode, contrary to what happens in the
massless case. 
In the expansion for $\theta$ we find the $\theta_{(i)}$ coefficients that
contain both the non-normalizable $i=0$ and the normalizable $i=4$ mode. The
$\Psi,\,\tilde\Psi$ terms appear due to the coupling to the gauge field.\\  From
the boundary term of the O.S. action one can obtain the 1-point function of the
dual operator $J^i$.

 As usual, it is convenient to group all the fields in a vector of the
appropriately normalized fields\footnote{Since the gauge field diverges at the
boundary precisely as $\sim r^\Delta$, this choice for the normalization allows
us to have a finite BBP and to collect the sources of the dual theory in
$\psi_{(0)}$. } $\psi=( r^{-\Delta} a_i, \phi$)  and express them as the (matrix
valued) bulk to boundary propagator (BBP) times a vector $\psi_{(0)}$ made of
the value of the sources.

\begin{equation}
\psi_I(r)= F_{IJ}(r) \psi_{J(0)} .
   \hspace{2cm}
F(\Lambda)= \mathbb{I}
 \end{equation}

moreover, it will be useful to separate the BBP matrix in a rectangular matrix
$\mathbb{F}$ and a vector $\mathbb{G}$ such that

\begin{equation}\label{partition}
 F= \left(
\begin{BMAT}[2pt]{ccc}{ccc:c} 
&&\\
&\mathbb{F}&\\
&&\\
 & \mathbb{G}&
\end{BMAT} 
\right)
 \hspace{2cm} a_I= r^\Delta \mathbb{F}_{IJ} \psi_{J(0)}  \hspace{2cm}   \phi = 
\mathbb{G}_{J} \psi_{J(0)}.  
\end{equation} 

In terms of these the expectation value of the current reads 
\begin{align}\label{current}
\langle J^m \rangle= &\lim_{r\rightarrow \infty} \,\,\sqrt[]{-g} \left( r^\Delta
\mathbb{F}_{im}  \left(F^{ir}+\frac{4\kappa}{3} \epsilon^{ijkl} \left(
A_j-\partial_j \theta \right)F_{kl} \right)     -\mathbb{G}_m(
F_{ij}F_{kl}\epsilon^{rijkl}+m^2(A^r-\partial^r\theta)  \right)+ \nonumber\\
& \lim_{r\rightarrow \infty} \,\,\sqrt[]{-\gamma} \, r^\Delta \mathbb{F}_{im}
\left (    \Delta (A^i-\partial^i\theta) + \frac{1}{2(\Delta+2)}   \partial^i
(\partial_j A^j - \Box \theta) -\frac{1}{2\Delta} \partial_j F^{j i}   \right) 
+ \nonumber\\
&\lim_{r\rightarrow \infty}\,\, \sqrt[]{-\gamma} \,\,\,\,\,\,\,\,\mathbb{G}_m 
\left(   \Delta  (\partial_j A^j - \Box \theta)   + \frac{1}{2(\Delta+2)} \Box
(\partial_j A^j - \Box \theta)    \right).   
\end{align}

The above expression is quite messy and needs some inspection. In the massless
case \cite{Gynther:2010ed} all terms proportional to $\mathbb{F}_{i\neq
m},\mathbb{G}_m $ vanish in the $r\rightarrow \infty$ limit and therefore are
not explicitly written in the literature. When the mass is present, however, all
terms in the expression are divergent. This is easy to check given the
expansions (\ref{expansion}). To have a better understanding of the properties
of the current it is convenient to collect the terms that do not contain finite
contributions as shown in the main text (\ref{currentx}).

\subsection{2-point functions}\label{app:2pfu1}

Equation (\ref{mirala}) is the correct expression for the correlator $\langle
J_y J_z \rangle$. However, one usually does not have an analytic solution for
the e.o.m. and therefore one has to construct the BBP numerically. This implies
that we are interested in (\ref{mirala}) expressed as a linear combination of
the BBP and its derivatives. In principle one can derive this combination
directly from the O.S. action to second order in perturbations but this might be
rather tedious. A simpler strategy is to look at the asymptotic expansions for
the perturbations and then invert the series to find the expression of the
normalizable mode as a combination of $\mathbb{F},\dot{\mathbb{F}}$.  The
anomalous conductivity (\ref{eq:s55}) is a good opportunity to perform an
explicit example.\\

First we switch on perturbations for all fields with momentum $k$ aligned to the
x direction and frequency $\omega$:  $\delta \theta = \sigma(r)e^{-i\omega t +
ikx}$ and $\delta A_{\mu} = a_{\mu}(r)e^{-i\omega t + ikx}$. The linearized
e.o.m. for these perturbations naturally separate in decoupled sectors, since we
are interested in the correlator $\langle J_y J_z  \rangle$ we just look at
\begin{align}
%f \sigma'' + \left(f' +\frac{3f}{r} \right) \sigma'+
%\left(\frac{\omega^2}{f}-\frac{k^2}{r^2}\right) \sigma - \frac{i \omega}{f}a_t -
%\frac{i k}{r^2}a_x =0\\
%f a_t'' + \frac{3f}{r}a_t' +\left(-2m^2- \frac{k^2}{r^2}\right)a_t
%-\frac{\omega k}{r^2}a_x - 2 i m^2 \omega \sigma = 0\\
%f a_x'' + \left( f' + \frac{f}{r}\right)a_x' + \left(-2m^2 +
%\frac{\omega^2}{f}\right)a_x + \frac{\omega k}{f}a_t +  2im ^2 k \sigma = 0 \\
f a_y'' + \left( f' + \frac{f}{r}\right)a_y' + \left(-m^2 + \frac{\omega^2}{f} -
\frac{k^2}{r^2}\right)a_y -\frac{8 i k \kappa \phi' a_z }{r} = 0 \label{eq:2}\\
f a_z'' + \left( f' + \frac{f}{r}\right)a_z' + \left(-m^2 + \frac{\omega^2}{f} -
\frac{k^2}{r^2}\right)a_z+\frac{8 i k \kappa \phi' a_y }{r} = 0 \label{eq:3}
\end{align}
that decouple from the other equations.

The asymptotic analysis of equations (\ref{eq:2},\ref{eq:3}) reveals that close
to the boundary the perturbations behave as
\begin{align}\label{expansion2}
a_i(r\rightarrow \infty)\sim a_{(0)i}\left(r^\Delta - \frac{k^2}{4 \Delta}
r^{\Delta-2} \right)+ a_{(0)j}\epsilon_{ij} \frac{8 \mu k \kappa
i}{3(\Delta-2)}r^{2\Delta-2} + \frac{\tilde a_i}{r^{2+\Delta}}.
\end{align}

Where $\tilde a_i$ is the normalizable mode of the perturbation. In principle it
has a complicated dependence on the sources but in the linear response regime we
can write
\begin{align}
\tilde a_i= \rho a_{i(0)} + \tilde \rho a_{j(0)} \longrightarrow \frac{\delta
\tilde a_{i}}{\delta a_{(0)j}} = \tilde \rho.
\end{align} 
that allows us to write (\ref{expansion2}) as
\begin{equation}\label{expansion3}
a_i(r\rightarrow \infty)\sim a_{(0)i}\left(r^\Delta - \frac{k^2}{4 \Delta}
r^{\Delta-2}+ \frac{\rho}{r^{\Delta+2}} \right)+ a_{(0)j}\epsilon_{ij} \left(
\frac{8 \mu k \kappa i}{3(\Delta-2)}r^{2\Delta-2} + \frac{\tilde
\rho}{r^{2+\Delta}} \right).
\end{equation}
Which is more useful to make the connection to the BBP matrix 
\begin{align}
\mathbb{F}= \begin{pmatrix} b(r)&c_+(r)\\c_-(r)&d(r)  \end{pmatrix}       
\end{align}
with \footnote{Here we make some abuse of language when we refer the block in
$\mathbb{F}$ that affects $a_x, \,a_y$ as $\mathbb{F}$. The true $\mathbb{F}$ is
actually a $4\times5$ matrix as explained in (\ref{partition}).}  
\begin{equation}
 b(r)=d(r)\sim 1 - \frac{k^2}{4 \Delta r^{2}} + \frac{\rho }{r^{2+2\Delta}}
\hspace{1cm}  c_{\pm} \sim \pm \left(\frac{8 \mu k \kappa
i}{3(\Delta-2)}r^{\Delta-2} + \frac{\tilde\rho }{r^{2+2\Delta}} \right).
\end{equation}
 At this point we can invert the series to the order of the normalizable mode.
In our concrete case we have
\begin{equation}
\tilde \rho=\lim_{r\rightarrow\infty}
r^{2+2\Delta}\frac{(2-\Delta)c(r)+rc'(r)}{-3\Delta} .
\end{equation}

 So the last thing to do is to numerically construct the BBP imposing infalling
boundary conditions at the horizon and compute the latter formula. For a
detailed explanation on how to numerically construct the BBP we refer the reader
to \cite{Kaminski:2009dh}. Due to how we numerically construct $\mathbb{F}$, one
may find some issues when computing $\lim_{r\rightarrow \infty}c(r)$ so we
rather use an alternative expression involving only derivatives of $c(r)$. One
can easily derive
\begin{equation}
\tilde \rho=\lim_{r\rightarrow\infty}
r^{3+2\Delta}\frac{(3-\Delta)c'(r)+rc''(r)}{6\Delta(\Delta+1)} .
\end{equation}
This expression combined with equations (\ref{mirala},\ref{eq:s55}) leads
finally to a expression for the conductivity
\begin{equation}
\sigma_{55}=\left. \lim_{k\rightarrow 0} \frac{i}{k_x} \lim_{r\rightarrow\infty}
r^{3+2\Delta}\frac{(3-\Delta)c'(r)+rc''(r)}{6\Delta}  \right|_{\omega=0}.
\end{equation}

%%%%%%%%%%%%%%%%%%%%%%%%%%%%%%%%%%%%%%%%%%%%%%%%%%%%%%%%%%%%%%%%%%%%%%%%%%%%%%%%
%%%%%%%%%%%%%%%%%%%%%%%%%%%%%%%%%%%%%%%%%%%%%%%%%%%%%%%%%%%%%%%%%%%%%%%%%%%%%%%%
%%%%%%%%%%%%%%%%%%%%%%%%%%%%%%%%%%%%%%%%
%%%%%%%%%%%%%%%%%%%%%%%%%%%%%%%%%%%%%%%%%%%%%%%%%%%%%%%%%%%%%%%%%%%%%%%%%%%%%%%%
%%%%%%%%%%%%%%%%%%%%%%%%%%%%%%%%%%%%%%%%%%%%%%%%%%%%%%%%%%%%%%%%%%%%%%%%%%%%%%%%
%%%%%%%%%%%%%%%%%%%%%%%%%%%%%%%%%%%%%%%%
%%%%%%%%%%%%%%%%%%%%%%%%%%%%%%%%%%%%%%%%%%%%%%%%%%%%%%%%%%%%%%%%%%%%%%%%%%%%%%%%
%%%%%%%%%%%%%%%%%%%%%%%%%%%%%%%%%%%%%%%%%%%%%%%%%%%%%%%%%%%%%%%%%%%%%%%%%%%%%%%%
%%%%%%%%%%%%%%%%%%%%%%%%%%%%%%%%%%%%%%%%
                                                                                
                          %%%%%%%%%%%%%%%%%%%%%%%%%%% 
%%%%%%%%%%%%%%%%%%%%%%%%%%%%%%%%%%%%%%%%%%%%%%%%%%            U1xU1 n-point
functions      %%%%%%%%%%%%%%%%%%%%%%%%%%%%%%%%%%%%%%%%
                                                                                
                          %%%%%%%%%%%%%%%%%%%%%%%%%%%
%%%%%%%%%%%%%%%%%%%%%%%%%%%%%%%%%%%%%%%%%%%%%%%%%%%%%%%%%%%%%%%%%%%%%%%%%%%%%%%%
%%%%%%%%%%%%%%%%%%%%%%%%%%%%%%%%%%%%%%%%%%%%%%%%%%%%%%%%%%%%%%%%%%%%%%%%%%%%%%%%
%%%%%%%%%%%%%%%%%%%%%%%%%%%%%%%%%%%%%%%%
%%%%%%%%%%%%%%%%%%%%%%%%%%%%%%%%%%%%%%%%%%%%%%%%%%%%%%%%%%%%%%%%%%%%%%%%%%%%%%%%
%%%%%%%%%%%%%%%%%%%%%%%%%%%%%%%%%%%%%%%%%%%%%%%%%%%%%%%%%%%%%%%%%%%%%%%%%%%%%%%%
%%%%%%%%%%%%%%%%%%%%%%%%%%%%%%%%%%%%%%%%
%%%%%%%%%%%%%%%%%%%%%%%%%%%%%%%%%%%%%%%%%%%%%%%%%%%%%%%%%%%%%%%%%%%%%%%%%%%%%%%%
%%%%%%%%%%%%%%%%%%%%%%%%%%%%%%%%%%%%%%%%%%%%%%%%%%%%%%%%%%%%%%%%%%%%%%%%%%%%%%%%
%%%%%%%%%%%%%%%%%%%%%%%%%%%%%%%%%%%%%%%%

\section{Correlators in the U(1)xU(1) model}

\subsection{1-point functions}\label{app:1pfu1u1}

 First of all we expand the action to first order in perturbations
\begin{align}\label{1order2}
 S^{(1)}_{\mathcal{R}}= &\int dr\,d^4x\,\, \sqrt[]{-g} \left[  a_\mu
\left(\nabla_\nu F^{\nu \mu} -m^2 (A^\mu-\partial^\mu \theta)+ \frac{3\kappa}{2}
\epsilon^{\mu \alpha \beta \gamma \rho} (F_{\alpha \beta} F_{\gamma
\rho}+H_{\alpha \beta} H_{\gamma \rho})   \right)\right] +\nonumber\\
&\int dr\,d^4x\,\, \sqrt[]{-g} \left[  v_\mu \left(\nabla_\nu H^{\nu \mu}+ 3
\kappa \epsilon^{\mu \alpha \beta \gamma \rho} F_{\alpha \beta} H_{\gamma \rho} 
 \right)-\phi  \nabla_\mu \left( A^\mu -\partial^\mu \theta   \right)  \right]
+\nonumber\\
& \int_\partial d^4x  \,\, \sqrt[]{-g} \left[ \,a_i \left ( F^{i r}+ 2\kappa
(A_j-\partial_j \theta) F_{kl} \epsilon^{r i j k l} \right)  \right]+\nonumber\\
& \int_\partial d^4x  \,\, \sqrt[]{-g} \left[ \,v_i \left ( H^{i r}+6 \kappa
(A_j-\partial_j \theta) H_{kl} \epsilon^{r i j k l} \right)  - \phi (F_{ij}
F_{kl} \epsilon^{r i j k l}+m^2(A^r-\partial^r\theta)) \right]+\nonumber\\
& \int_\partial d^4x  \,\, \sqrt[]{-\gamma} \,\,a_i \left (    \Delta
(A^i-\partial^i\theta) + \frac{1}{2(\Delta+2)}   \partial^i (\partial_j A^j -
\Box \theta) -\frac{1}{2\Delta} \partial_j F^{j i}   \right)  + \nonumber\\
& \int_\partial d^4x  \,\, \sqrt[]{-\gamma} \,\,v_i   \left(    -\frac{1}{2}
\partial_j H^{j i} \log(r) \right)   \nonumber\\
& \int_\partial d^4x  \,\, \sqrt[]{-\gamma} \,\,\phi    \left(   \Delta 
(\partial_j A^j - \Box \theta)   + \frac{1}{2(\Delta+2)} \Box (\partial_j A^j -
\Box \theta)    \right) . 
\end{align}

From the e.o.m. we find that the expansions for the scalar and the massive gauge
field remain qualitatively unchanged up to the normalizable mode w.r.t. what we
found in the $U(1)$ model. The expansion for the vector field is
\begin{equation}
V_\mu=\sum_i V_{\mu(i)}r^{-i} + \sum_{i\geq2} \tilde V_{\mu(i)}r^{-i}\log(r) +
\sum_{n,i\geq n+1} \Lambda_{\mu(n,i)}    r^{n\Delta-i}
\end{equation}

Where the $\sim \Lambda$ terms appear due to the mixing with the axial gauge
field via Chern Simons. As in the previous case it is convenient to define the
BBP with the fields normalized $(r^{-\Delta} a_i,v_i, \phi) $ so that we can
impose
 \begin{equation}
\psi_{I}(0)\equiv \begin{pmatrix} a_{t(0)}\\\vdots\\v_{t(0)}\\\vdots
\\\phi_{(0)} \end{pmatrix}    \hspace{2cm}
F(\Lambda)= \mathbb{I}.
 \end{equation}
I is useful to divide the BBP in two rectangular matrices $\mathbb{F},
\mathbb{H}$ and a vector $\mathbb{G}$
\begin{equation}
 F= \left(
\begin{BMAT}[3pt]{ccccc}{ccc:ccc:c} 
&&&&\\
&&\mathbb{F}&&\\
&&&&\\
&&&&\\
&&\mathbb{H}&&\\
&&&&\\
 && \mathbb{G}&&
\end{BMAT} 
\right)
 \hspace{1cm} a_I= r^\Delta \mathbb{F}_{IJ} \psi_{J(0)}  \hspace{1cm}  v_I= 
\mathbb{H}_{IJ} \psi_{J(0)}  \hspace{1cm}  \phi =  \mathbb{G}_{J} \psi_{J(0)}.  
\end{equation} 

From this one can derive the renormalized 1-point functions. The expressions can
be found in the main text in (\ref{correintev}).

\subsection{2-point functions}
\label{app:2pfu1u1}

In order to obtain the 2-point functions in
(\ref{kubo},\ref{kubos},\ref{kuboss}) we switch on perturbations with momentum
aligned to the $z$ direction $\delta \theta = \sigma(r)e^{-i\omega t + ikz}$,
$\delta A_{\mu} = a_{\mu}(r)e^{-i\omega t + ikz}$ and $\delta V_{\mu} =
v_{\mu}(r)e^{-i\omega t + ikz}$ on top of our background (\ref{bakbakbak}). The
equations  decouple and in the sector we are interested in we are left to four
coupled equations for $a_x, a_y, v_x, v_y$. 
\begin{align}
a_y''+\left( \frac{f'}{f}+\frac{1}{r}\right)a_y' + \left(
\frac{\omega^2}{f^2}-\frac{k^2 }{r^2 f}-\frac{m^2}{f}\right)a_y- \frac{12 i k
\kappa \phi'}{f r}a_z- \frac{12 i k \kappa \chi'}{f r}v_z =0\\
a_z''+\left( \frac{f'}{f}+\frac{1}{r}\right)a_z' + \left(
\frac{\omega^2}{f^2}-\frac{k^2 }{r^2 f}-\frac{m^2}{f}\right)a_z+ \frac{12 i k
\kappa \phi'}{f r}a_y + \frac{12 i k \kappa \chi'}{f r}v_y =0\\
v_y''+\left( \frac{f'}{f}+\frac{1}{r}\right)v_y' +
\left(\frac{\omega^2}{f^2}-\frac{k^2 }{r^2 f}\right)v_y- \frac{12 i k \kappa
\chi'}{f r}a_z - \frac{12 i k \kappa \phi'}{f r}v_z =0\\
v_z''+\left( \frac{f'}{f}+\frac{1}{r}\right)v_z' +
\left(\frac{\omega^2}{f^2}-\frac{k^2 }{r^2 f}\right)v_z+ \frac{12 i k \kappa
\chi'}{f r}a_y + \frac{12 i k \kappa \phi'}{f r}v_y =0
\end{align}

The asymptotic analysis of these equations allows to write the near boundary
expansion 
\begin{align}
&a_i(r\rightarrow \infty)\sim a_{(0)i}(r^\Delta + M r^{\Delta-2})+
a_{(0)j}\epsilon_{ij}\tilde M r^{2\Delta-2} + \frac{\tilde{a_i}}{r^{\Delta+2}}\\
&v_i(r\rightarrow \infty)\sim v_{(0)i}(1 )+ v_{(0)j}\epsilon_{ij}( \tilde M
r^{\Delta-2}) + \frac{\tilde{v_i}}{r^{2}}.
\end{align}
Where $M$ and $\tilde M$ are functions of $k,\kappa, A_t', V_t '$. In the linear
response limit the normalizable modes $\tilde a_ i\, \tilde v_i$ can only depend
linearly on the sources, therefore we may rewrite the expansions 
\begin{align}
&a_i(r\rightarrow \infty)\sim a_{(0)i}(r^\Delta + M r^{\Delta-2} + \frac{\rho
}{r^{2+\Delta}})+ a_{(0)j}\epsilon_{ij}( \tilde M r^{2\Delta-2} +
\frac{\tilde\rho }{r^{2+\Delta}})+ v_{(0)i}
\frac{\tilde{\tilde\rho}}{r^{2+\Delta}}  +  v_{(0)j}\epsilon_{ij}
\frac{\tilde{\tilde{\tilde\rho}} }{r^{2+\Delta}}\\
&v_i(r\rightarrow \infty)\sim v_{(0)i}(1 + \frac{\eta }{r^{2}})+
v_{(0)j}\epsilon_{ij}( \tilde M r^{\Delta-2} + \frac{\tilde\eta }{r^{2}})+   
a_{(0)i} \frac{\tilde{\tilde\eta} }{r^{2}}    +    a_{(0)j}\epsilon_{ij}
\frac{\tilde{\tilde{\tilde\eta}} }{r^{2}}.  
\end{align}
This allows us to write
\begin{align}
&\langle J_i^V J_j^V\rangle= 2 \frac{\delta \tilde v_i}{\delta v_{j(0)}}=
2\tilde \eta_i\\
&\langle J_i^A J_j^A\rangle= (2+2\Delta) \frac{\delta \tilde a_i}{\delta
a_{j(0)}}= (2+2\Delta)\tilde \rho_i\\
&\langle J_i^A J_j^V\rangle= 2 \frac{\delta \tilde v_i}{\delta a_{j(0)}} =
2\tilde{\tilde{\tilde \eta}}_i.
\end{align}

Now we perform the same analysis as in (\ref{app:2pfu1}), seeking the correct
expression of these correlators as a linear combination of the BBP and its
derivatives in order to compute the conductivities numerically. We find
\begin{align}
2\tilde \eta_i&=\lim_{r\rightarrow\infty} -r^{3} p'(r)\\
 (2+2\Delta)\tilde \rho_i&=\lim_{r\rightarrow\infty}
r^{3+2\Delta}\frac{(3-\Delta)b'(r)+rb''(r)}{3\Delta} \\
 2\tilde{\tilde{\tilde \eta}}_i&=\lim_{r\rightarrow\infty}
-r^{3+2\Delta}\frac{v'(r)}{\Delta+1} 
\end{align}
where $p(r)$, $ b(r)$ and $v(r)$ are the functions that appear in the matrix
valued BBP 
\begin{align}
 \begin{pmatrix}
r^{-\Delta}a_y(r)\\
r^{-\Delta}a_z(r)\\
v_y(r)\\
v_z(r)
 \end{pmatrix}=
 \begin{pmatrix}
 a(r)&b(r)&&c(r)&&&d(r)\\
 i(r)&j(r)&&k(r)&&&l(r)\\
 m(r)&n(r)&&o(r)&&&p(r)\\
u(r)&v(r)&&w(r)&&&y(r)
 \end{pmatrix} \begin{pmatrix}
a_{y(0)}\\
a_{z(0)}\\
v_{y(0)}\\
v_{z(0)}
 \end{pmatrix}\end{align}

\section{U(1)xU(1) Model: perturbations for the CMW}\label{app:eomcmw}

In order to compute the QNM spectrum and the electric conductivities with a
constant and homogeneous background magnetic field we switch on perturbations
with momentum $k$ aligned to the magnetic field and frequency $\omega$. The
decoupled sector of equations we are interested in reads
\begin{align}
a_t'' + \frac{3}{r}at' - \left( \frac{k^2}{f r^2}+\frac{m^2}{f}\right)a_t
-\frac{\omega k}{f r^2}a_z+\frac{12 \kappa B}{r^3}v_z' +\frac{i \omega
m^2}{f}\eta=0\\
v_t'' + \frac{3}{r}vt' -\frac{k^2}{f r^2}v_t -  \frac{\omega k}{f
r^2}v_z+\frac{12 \kappa B}{r^3}a_z' =0\\
a_z''+\left( \frac{f'}{f}+\frac{1}{r}\right)a_z' + \left(
\frac{\omega^2}{f^2}-\frac{m^2}{f}\right)a_z+\frac{\omega k}{f^2}a_t+\frac{12
\kappa B}{f r}v_t'-\frac{i k m^2}{f}\eta=0\\
v_z''+\left( \frac{f'}{f}+\frac{1}{r}\right)v_z' +
\frac{\omega^2}{f^2}v_z+\frac{\omega k}{f^2}v_t+\frac{12 \kappa B}{f r}a_t'=0\\
\eta'' + \left(\frac{3}{r} +
\frac{f'}{f}\right)\eta'+\left(\frac{\omega^2}{f^2}- \frac{k^2}{f} \right)\eta
+\frac{i \omega}{f^2}a_t + \frac{i k}{f r}a_z=0.
\end{align}

With $a$, $v$, $\eta$ being the perturbations for the axial, vector and
St\"uckelberg fields respectively and f the blackening factor of the metric.
There are as well two constraints:
\begin{align}
\omega a_t' + \frac{k f}{r^2}a_z' +\frac{12\kappa B}{r^3}\left( \omega v_z+k
v_t\right)-i m^2 f \eta '=0\\
\omega v_t' + \frac{k f}{r^2}v_z' +\frac{12\kappa B}{r^3}\left( \omega a_z+k
a_t\right)=0.
\end{align}
\\

The equations for the electric conductivity can be obtained turning off the
momentum.

\end{appendix}

\newpage

\addcontentsline{toc}{section}{References}

\nocite{*}
\bibliographystyle{jhepcap}
\bibliography{Sb}

\providecommand{\href}[2]{#2}\begingroup\raggedright\begin{thebibliography}{10}

\bibitem{Adler:1969gk}
S.~L. Adler, {\it {Axial vector vertex in spinor electrodynamics}},  {\em
  Phys.Rev.} {\bf 177} (1969) 2426--2438.

\bibitem{Bell:1969ts}
J.~Bell and R.~Jackiw, {\it {A PCAC puzzle: pi0 -> gamma gamma in the sigma
  model}},  {\em Nuovo Cim.} {\bf A60} (1969) 47--61.

\bibitem{Bertlmann:1996xk}
R.~A. Bertlmann, {\it {Anomalies in quantum field theory}}, . Oxford, UK:
  Clarendon (1996) 566 p. (International series of monographs on physics: 91).

\bibitem{Bilal:2008qx}
A.~Bilal, {\it {Lectures on Anomalies}},
  \href{http://xxx.lanl.gov/abs/0802.0634}{{\tt 0802.0634}}.

\bibitem{Fukushima:2008xe}
K.~Fukushima, D.~E. Kharzeev, and H.~J. Warringa, {\it {The Chiral Magnetic
  Effect}},  {\em Phys. Rev.} {\bf D78} (2008) 074033,
  [\href{http://xxx.lanl.gov/abs/0808.3382}{{\tt 0808.3382}}].

\bibitem{Erdmenger:2008rm}
J.~Erdmenger, M.~Haack, M.~Kaminski, and A.~Yarom, {\it {Fluid dynamics of
  R-charged black holes}},  {\em JHEP} {\bf 01} (2009) 055,
  [\href{http://xxx.lanl.gov/abs/0809.2488}{{\tt 0809.2488}}].

\bibitem{Banerjee:2008th}
N.~Banerjee {\em et.~al.}, {\it {Hydrodynamics from charged black branes}},
  {\em JHEP} {\bf 01} (2011) 094,
  [\href{http://xxx.lanl.gov/abs/0809.2596}{{\tt 0809.2596}}].

\bibitem{Kharzeev:2009pj}
D.~E. Kharzeev and H.~J. Warringa, {\it {Chiral Magnetic conductivity}},  {\em
  Phys. Rev.} {\bf D80} (2009) 034028,
  [\href{http://xxx.lanl.gov/abs/0907.5007}{{\tt 0907.5007}}].

\bibitem{Landsteiner:2011cp}
K.~Landsteiner, E.~Megias, and F.~Pena-Benitez, {\it {Gravitational Anomaly and
  Transport}},  {\em Phys. Rev. Lett.} {\bf 107} (2011) 021601,
  [\href{http://xxx.lanl.gov/abs/1103.5006}{{\tt 1103.5006}}].

\bibitem{Loganayagam:2012pz}
R.~Loganayagam and P.~Surowka, {\it {Anomaly/Transport in an Ideal Weyl gas}},
  {\em JHEP} {\bf 1204} (2012) 097,
  [\href{http://xxx.lanl.gov/abs/1201.2812}{{\tt 1201.2812}}].

\bibitem{Newman:2005hd}
G.~M. Newman, {\it {Anomalous hydrodynamics}},  {\em JHEP} {\bf 01} (2006) 158,
  [\href{http://xxx.lanl.gov/abs/hep-ph/0511236}{{\tt hep-ph/0511236}}].

\bibitem{Landsteiner:2011iq}
K.~Landsteiner, E.~Megias, L.~Melgar, and F.~Pena-Benitez, {\it {Holographic
  Gravitational Anomaly and Chiral Vortical Effect}},  {\em JHEP} {\bf 1109}
  (2011) 121, [\href{http://xxx.lanl.gov/abs/1107.0368}{{\tt 1107.0368}}].

\bibitem{Son:2009tf}
D.~T. Son and P.~Surowka, {\it {Hydrodynamics with Triangle Anomalies}},  {\em
  Phys. Rev. Lett.} {\bf 103} (2009) 191601,
  [\href{http://xxx.lanl.gov/abs/0906.5044}{{\tt 0906.5044}}].

\bibitem{Neiman:2010zi}
Y.~Neiman and Y.~Oz, {\it {Relativistic Hydrodynamics with General Anomalous
  Charges}},  {\em JHEP} {\bf 03} (2011) 023,
  [\href{http://xxx.lanl.gov/abs/1011.5107}{{\tt 1011.5107}}].

\bibitem{Banerjee:2012iz}
N.~Banerjee, J.~Bhattacharya, S.~Bhattacharyya, S.~Jain, S.~Minwalla, {\em
  et.~al.}, {\it {Constraints on Fluid Dynamics from Equilibrium Partition
  Functions}},  {\em JHEP} {\bf 1209} (2012) 046,
  [\href{http://xxx.lanl.gov/abs/1203.3544}{{\tt 1203.3544}}].

\bibitem{Jensen:2012kj}
K.~Jensen, R.~Loganayagam, and A.~Yarom, {\it {Thermodynamics, gravitational
  anomalies and cones}},  {\em JHEP} {\bf 1302} (2013) 088,
  [\href{http://xxx.lanl.gov/abs/1207.5824}{{\tt 1207.5824}}].

\bibitem{Golkar:2012kb}
S.~Golkar and D.~T. Son, {\it {Non-Renormalization of the Chiral Vortical
  Effect Coefficient}},  \href{http://xxx.lanl.gov/abs/1207.5806}{{\tt
  1207.5806}}.

\bibitem{Hou:2012xg}
D.-F. Hou, H.~Liu, and H.-c. Ren, {\it {A Possible Higher Order Correction to
  the Vortical Conductivity in a Gauge Field Plasma}},  {\em Phys.Rev.} {\bf
  D86} (2012) 121703, [\href{http://xxx.lanl.gov/abs/1210.0969}{{\tt
  1210.0969}}].

\bibitem{Jensen:2013vta}
K.~Jensen, P.~Kovtun, and A.~Ritz, {\it {Chiral conductivities and effective
  field theory}},  {\em JHEP} {\bf 1310} (2013) 186,
  [\href{http://xxx.lanl.gov/abs/1307.3234}{{\tt 1307.3234}}].

\bibitem{'tHooft:1986nc}
G.~'t~Hooft, {\it {How Instantons Solve the U(1) Problem}},  {\em Phys.Rept.}
  {\bf 142} (1986) 357--387.

\bibitem{Adler:2004qt}
S.~L. Adler, {\it {Anomalies to all orders}},
  \href{http://xxx.lanl.gov/abs/hep-th/0405040}{{\tt hep-th/0405040}}.

\bibitem{Ioffe:2006ww}
B.~Ioffe, {\it {Axial anomaly: The Modern status}},  {\em Int.J.Mod.Phys.} {\bf
  A21} (2006) 6249--6266, [\href{http://xxx.lanl.gov/abs/hep-ph/0611026}{{\tt
  hep-ph/0611026}}].

\bibitem{Klebanov:2002gr}
I.~R. Klebanov, P.~Ouyang, and E.~Witten, {\it {A Gravity dual of the chiral
  anomaly}},  {\em Phys.Rev.} {\bf D65} (2002) 105007,
  [\href{http://xxx.lanl.gov/abs/hep-th/0202056}{{\tt hep-th/0202056}}].

\bibitem{Gursoy:2014ela}
U.~Gursoy and A.~Jansen, {\it {(Non)renormalization of Anomalous Conductivities
  and Holography}},  \href{http://xxx.lanl.gov/abs/1407.3282}{{\tt 1407.3282}}.

\bibitem{Landsteiner:2012kd}
K.~Landsteiner, E.~Megias, and F.~Pena-Benitez, {\it {Anomalous Transport from
  Kubo Formulae}},  {\em Lect.Notes Phys.} {\bf 871} (2013) 433--468,
  [\href{http://xxx.lanl.gov/abs/1207.5808}{{\tt 1207.5808}}].

\bibitem{Kharzeev:2010gd}
D.~E. Kharzeev and H.-U. Yee, {\it {Chiral Magnetic Wave}},  {\em Phys.Rev.}
  {\bf D83} (2011) 085007, [\href{http://xxx.lanl.gov/abs/1012.6026}{{\tt
  1012.6026}}].

\bibitem{Casero:2007ae}
R.~Casero, E.~Kiritsis, and A.~Paredes, {\it {Chiral symmetry breaking as open
  string tachyon condensation}},  {\em Nucl.Phys.} {\bf B787} (2007) 98--134,
  [\href{http://xxx.lanl.gov/abs/hep-th/0702155}{{\tt hep-th/0702155}}].

\bibitem{Hartnoll:2008kx}
S.~A. Hartnoll, C.~P. Herzog, and G.~T. Horowitz, {\it {Holographic
  Superconductors}},  {\em JHEP} {\bf 0812} (2008) 015,
  [\href{http://xxx.lanl.gov/abs/0810.1563}{{\tt 0810.1563}}].

\bibitem{Franco:2009yz}
S.~Franco, A.~Garcia-Garcia, and D.~Rodriguez-Gomez, {\it {A General class of
  holographic superconductors}},  {\em JHEP} {\bf 1004} (2010) 092,
  [\href{http://xxx.lanl.gov/abs/0906.1214}{{\tt 0906.1214}}].

\bibitem{Gynther:2010ed}
A.~Gynther, K.~Landsteiner, F.~Pena-Benitez, and A.~Rebhan, {\it {Holographic
  Anomalous Conductivities and the Chiral Magnetic Effect}},  {\em JHEP} {\bf
  02} (2011) 110, [\href{http://xxx.lanl.gov/abs/1005.2587}{{\tt 1005.2587}}].

\bibitem{Kaminski:2009dh}
M.~Kaminski, K.~Landsteiner, J.~Mas, J.~P. Shock, and J.~Tarrio, {\it
  {Holographic Operator Mixing and Quasinormal Modes on the Brane}},  {\em
  JHEP} {\bf 1002} (2010) 021, [\href{http://xxx.lanl.gov/abs/0911.3610}{{\tt
  0911.3610}}].

\bibitem{Landsteiner:2011tf}
K.~Landsteiner, E.~Megias, L.~Melgar, and F.~Pena-Benitez, {\it {Gravitational
  Anomaly and Hydrodynamics}},  {\em J.Phys.Conf.Ser.} {\bf 343} (2012) 012073,
  [\href{http://xxx.lanl.gov/abs/1111.2823}{{\tt 1111.2823}}].

\bibitem{Amado:2011zx}
I.~Amado, K.~Landsteiner, and F.~Pena-Benitez, {\it {Anomalous transport
  coefficients from Kubo formulas in Holography}},  {\em JHEP} {\bf 1105}
  (2011) 081, [\href{http://xxx.lanl.gov/abs/1102.4577}{{\tt 1102.4577}}].

\bibitem{Burnier:2011bf}
Y.~Burnier, D.~E. Kharzeev, J.~Liao, and H.-U. Yee, {\it {Chiral magnetic wave
  at finite baryon density and the electric quadrupole moment of quark-gluon
  plasma in heavy ion collisions}},  {\em Phys.Rev.Lett.} {\bf 107} (2011)
  052303, [\href{http://xxx.lanl.gov/abs/1103.1307}{{\tt 1103.1307}}].

\bibitem{Nielsen:1983rb}
H.~B. Nielsen and M.~Ninomiya, {\it {ADLER-BELL-JACKIW ANOMALY AND WEYL
  FERMIONS IN CRYSTAL}},  {\em Phys.Lett.} {\bf B130} (1983) 389.

\bibitem{Gorbar:2013dha}
E.~Gorbar, V.~Miransky, and I.~Shovkovy, {\it {Chiral anomaly, dimensional
  reduction, and magnetoresistivity of Weyl and Dirac semimetals}},  {\em
  Phys.Rev.} {\bf B89} (2014) 085126,
  [\href{http://xxx.lanl.gov/abs/1312.0027}{{\tt 1312.0027}}].

\bibitem{Son:2012bg}
D.~Son and B.~Spivak, {\it {Chiral Anomaly and Classical Negative
  Magnetoresistance of Weyl Metals}},  {\em Phys.Rev.} {\bf B88} (2013) 104412,
  [\href{http://xxx.lanl.gov/abs/1206.1627}{{\tt 1206.1627}}].

\bibitem{Nakamura:2009tf}
S.~Nakamura, H.~Ooguri, and C.-S. Park, {\it {Gravity Dual of Spatially
  Modulated Phase}},  {\em Phys.Rev.} {\bf D81} (2010) 044018,
  [\href{http://xxx.lanl.gov/abs/0911.0679}{{\tt 0911.0679}}].

\bibitem{Bhattacharyya:2008jc}
S.~Bhattacharyya, V.~E. Hubeny, S.~Minwalla, and M.~Rangamani, {\it {Nonlinear
  Fluid Dynamics from Gravity}},  {\em JHEP} {\bf 0802} (2008) 045,
  [\href{http://xxx.lanl.gov/abs/0712.2456}{{\tt 0712.2456}}].

\bibitem{Papadimitriou:2004ap}
I.~Papadimitriou and K.~Skenderis, {\it {AdS / CFT correspondence and
  geometry}},  \href{http://xxx.lanl.gov/abs/hep-th/0404176}{{\tt
  hep-th/0404176}}.

\end{thebibliography}\endgroup

\end{document}